\documentclass[a4paper,onecolumn,11pt]{quantumarticle}
\pdfoutput=1
\usepackage[utf8]{inputenc}
\usepackage[english]{babel}
\usepackage[T1]{fontenc}
\usepackage{amsmath}
\usepackage{hyperref}

\usepackage{tikz}
\usepackage{lipsum}

\usepackage{graphicx}
\usepackage{subcaption}
\usepackage[numbers]{natbib}
\usepackage{booktabs}
\usepackage{multirow}
\usepackage{amsmath}

\begin{document}

\title{Expressibility-Enhancing Strategies for Quantum Neural Networks}

\author{Yalin Liao}
\affiliation{University of Delaware, Newark, DE 19716, USA}
\affiliation{Alfred University, Alfred, NY 14802, USA}
\orcid{0000-0002-2445-2701}
\author{Junpeng Zhan}
\email{zhanj@alfred.edu, zhanjunpeng@gmail.com}
\homepage{https://sites.google.com/site/eejzhan/home}
\orcid{0000-0003-0290-4698}
\thanks{Yalin Liao contributed to this work when he was a visiting scholar at Alfred University in the Summer of 2022. He is a Ph.D. student at the University of Delaware. Both authors thank the National Science Foundation for its support under Award ERI 2138702.}
\affiliation{Alfred University, Alfred, NY 14802, USA}
\maketitle
\begin{abstract}
 Quantum neural networks (QNNs), represented by parameterized quantum circuits, can be trained in the paradigm of supervised learning to map input data to predictions. Much work has focused on theoretically analyzing the expressive power of QNNs. However, in almost all literature, QNNs' expressive power is numerically validated using only simple univariate functions. We surprisingly discover that state-of-the-art QNNs with strong expressive power can have poor performance in approximating even just a simple sinusoidal function. To fill the gap, we propose four expressibility-enhancing strategies for QNNs: Sinusoidal-friendly embedding, redundant measurement, post-measurement function, and random training data. We analyze the effectiveness of these strategies via mathematical analysis and/or numerical studies including learning complex sinusoidal-based functions. Our results from comparative experiments validate that the four strategies can significantly increase the QNNs' performance in approximating complex multivariable functions and reduce the quantum circuit depth and qubits required. 
\end{abstract}

\section{Introduction}
Quantum computing is the exploitation of collective properties of quantum states, such as superposition and entanglement, to perform computation via quantum computers \cite{nielsen_chuang_2010}. Quantum computers are developing rapidly. In 2019, the Google AI Quantum team achieved quantum supremacy on a 53-qubit processor over a state-of-the-art supercomputer \cite{Arute2019Nature_Google_quantum_supremacy}.

A universal fault-tolerant quantum computer with millions of qubits and long coherence times can be extremely powerful but will potentially take decades of research. The number of qubits largely determines the power of a quantum computer, and it is widely expected to be increasing fast, e.g., IBM plans to debut an 1121-qubit quantum processor in 2023 \footnote{\url{https://research.ibm.com/blog/ibm-quantum-roadmap-2025}}. \textit{Noisy intermediate-scale quantum (NISQ) computers} with 208-420 qubits, used together with classical computers, \textit{are expected to achieve sustainable quantum supremacy over classical computers} \cite{Bharti_2022NISQ, Preskill2018quantumcomputingin}. 

Deep learning, a subarea in machine learning (ML), has recently become incredibly powerful in various important fields. However, neural networks (NNs) used in deep learning can be extremely large, e.g., having $\sim$100 billion parameters. Training such NNs can be extremely challenging, time-consuming, and costly. To address this issue, quantum computing has been investigated to ease the training of NNs, although it is still in its infancy. Variational quantum algorithms (VQAs), as hybrid quantum and classical algorithms \cite{Peruzzo2014vqe}, are promising in realizing supremacy and have gained a lot of attention recently \cite{McArdle2019, Mitarai2020VQE, Li2017PRX}. The basic idea of the hybrid approach is to split a problem into two parts which are solved by classical and quantum computers, respectively. This paradigm could help achieve quantum computing supremacy in the short term. Quantum ML (QML) is one of the most important subareas of general VQAs and quantum neural network (QNN) is a very important branch of QML. \textit{Notably, QNNs can represent more complex functions than classical NNs \cite{mitarai2018QCL}.} Since expressive power is the core of QNN, we focus on enhancing QNNs' expressibility in this paper.

\paragraph{Motivation: }Much work has focused on analyzing the expressive power of QNNs \cite{mitarai2018QCL,schuld2021effect}. But existing QNNs' expressibility is usually numerically validated by univariate functions (we have not found any validation on multivariable functions in the literature). 
In addition, our experiments confirm that a QNN that has been claimed to have a strong expressibility can perform poorly in approximating even just a simple univariate function such as $\sin(2\pi x)$. This indicates that \textit{current QNNs are weak in complex learning tasks. Therefore, there is a need to realize QNNs that can practically approximate more complex multivariable functions}.

\begin{figure}[t]
    \centering
    \includegraphics[scale=0.4]{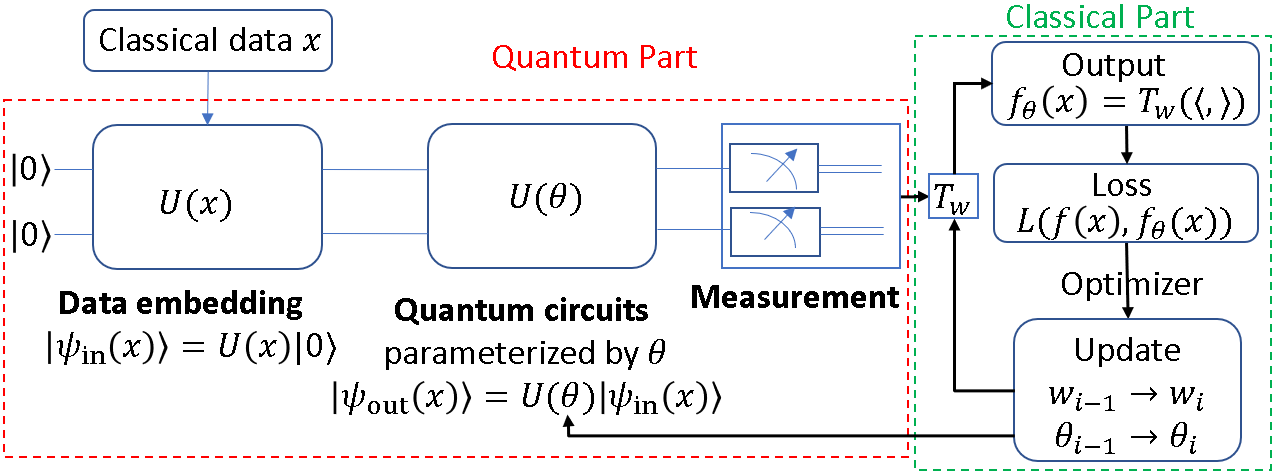}
    \caption{Schematic diagram of our QNN.}
    \label{figure_QNN_diagram}
\end{figure}

\subsection{Main Contributions}
This work has five contributions including four strategies to increase the expressive power of QNN which is depicted in Figure \ref{figure_QNN_diagram}: 
(i) We propose a new angle embedding that is most suitable for sinusoidal-based multivariable nonlinear target functions. 
(ii) We introduce a redundant measurement into the QNN. Our analysis and experiment studies show that this strategy introduces more parameters to QNNs, making them easier to be trained without additional quantum resources such as quantum circuit depth and gates. 
(iii) We propose to apply a post-measurement function ($T_w$ in Figure \ref{figure_QNN_diagram}) to quantum measurement output. Our analysis shows that this strategy can increase the number of different basis functions, hence enhancing the QNNs' expressive power without using more quantum resources. 
(iv) We find that using random training samples instead of using uniformly spaced samples also improves QNNs' performance. 
(v) Using these four strategies, we obtain QNNs that accurately approximate sinusoidal-based multivariable functions, which have not been seen in the existing literature. The first three strategies can also help to enhance QNNs' performance in other ML tasks such as classification.

\subsection{Related Work}
As noted in \cite{schuld2021effect}, there are different names for QNNs \cite{McClean2018Barren, Farhi2018ClassificationWQ}, e.g., variational hybrid quantum-classical algorithms \cite{McClean_2016}, quantum circuit learning \cite{mitarai2018QCL}, parameterized quantum circuits \cite{Benedetti_2019}, variational quantum ML \cite{schuld2021effect}, etc.

Quantum algorithms \cite{Montanaro2016} can be classified into three families: gate circuits (e.g., Grover’s and Shor’s algorithms \cite{nielsen_chuang_2010}), annealing \cite{Brady2021PRL_quantum_annealing}, and VQAs \cite{Peruzzo2014vqe,McClean_2016}. As indicated in \cite{Garcia2022QMLreview}, the main types of QML algorithms are quantum implementations of classical ML. QNNs, as a kind of general VQAs, are a very active research area in QML, thanks in part to the great success of deep learning. Different types of QNNs have been developed, e.g., quantum convolutional NN \cite{Li_2020} and quantum generative adversarial network \cite{Anand_2021, Niu2021GAN}. QNNs have also been used in different fields such as classification \cite{Silver_AAAI2022QUILT, Heidari_Grama_Szpankowski_2022}, regression \cite{Paquet2022}, and natural language processing \cite{Coecke2020}. Several nice reviews of QML, including QNNs, are given in \cite{Garcia2022QMLreview,Biamonte2017Nat, Dunjko_2018}.

Existing research on QNNs for supervised learning has covered several different aspects: parameter optimization methods \cite{Stokes2020quantumnatural, Kubler2020adaptiveoptimizer}, structure optimization methods \cite{Ostaszewski2021structure, Du2022}, objective function setting such as how to avoid barren plateau \cite{Cerezo2021NatComBarren, McClean2018Barren}, the number of measurements \cite{Kubler2020adaptiveoptimizer}. In this paper, we focus on other components in QNNs that have been ignored or under-investigated in existing literature, i.e., use a redundant measurement and a post-measurement function to increase QNNs' expressive power. \cite{Kubler2020adaptiveoptimizer} argues that the number of measurements should be decreased. However, we discover that increasing the number of measurements can enhance QNNs' performance without using additional quantum resources.

Embedding methods play a vital role in the expressive power of QNNs. Papers \cite{schuld2021effect, Vidal2019} propose to repeat simple data encoding gates multiple times to increase the expressive power of QNNs. These two papers show their effectiveness in approximating univariate sinusoidal functions. Our first strategy is also about embedding but the focus is different: they focus on representing the different spectrums of sinusoidal functions while we focus on representing complex multivariable functions with and without sinusoidal components.
\cite{mitarai2018QCL} proposes a quantum circuit learning (QCL) that theoretically can approximate any analytical functions. However, we find that its practical fitting performance can significantly deteriorate when the target function becomes more complex. Our paper's sinusoidal-friendly embedding method is more suitable for expressing sinusoidal-based functions and is different from that used in \cite{mitarai2018QCL}. Other related topics include deep NNs for approximating functions \cite{Liang2016DNN} and expressive power analysis \cite{Du_2020}.

\section{Method}
As shown in Figure \ref{figure_QNN_diagram}, our QNN consists of data embedding, parameterized quantum circuit, measurement, a post-measurement function $T_w$, and an optimizer used to update parameters. This section proposes four strategies to enhance the QNN's expressibility.

\subsection{Strategy 1: Embedding}
\label{method_strategy1_embedding}
Embedding data into a quantum computer is a key step of quantum computing and can be done via several methods such as basic embedding, angle embedding, and amplitude embedding. The three embedding methods given below are different ways of using angle embedding. \cite{Schuld2021kernel} has more details about different embedding methods. 

\paragraph{Arcsin Embedding.} When the input to an angle embedding is $\sin^{-1}(x)$, we call it Arcsin embedding for the convenience of expression. For a one-qubit circuit, this embedding leads to basis functions given in $\mathcal{BF}_A=\left\{1, x, \sqrt{1-x^2}\right\}$, as shown in \cite{mitarai2018QCL}. Quantum circuits adopting this embedding could approximate any differentiable functions if enough qubits are utilized \cite{mitarai2018QCL}.
\begin{figure}[t]
    \centering
    \begin{subfigure}[b]{0.45\textwidth}
    \centering
    \includegraphics[width=0.92\textwidth]{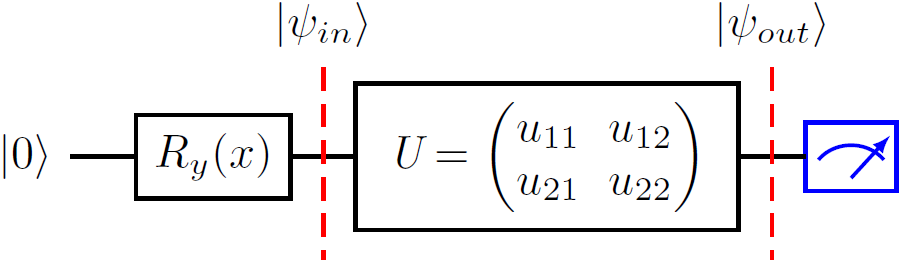}
    \caption{1-qubit, sinusoidal-friendly embedding}
    \end{subfigure}
    \begin{subfigure}[b]{0.45\textwidth}
    \centering
    \includegraphics[width=0.68\textwidth]{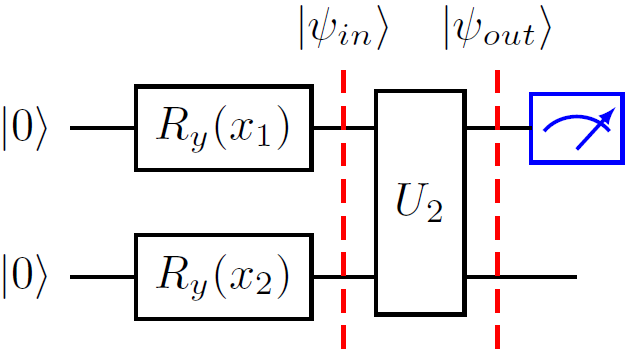}
    \caption{2-qubit, sinusoidal-friendly embedding}
    \end{subfigure}
    \begin{subfigure}[b]{0.45\textwidth}
    \centering
    \includegraphics[width=0.9\textwidth]{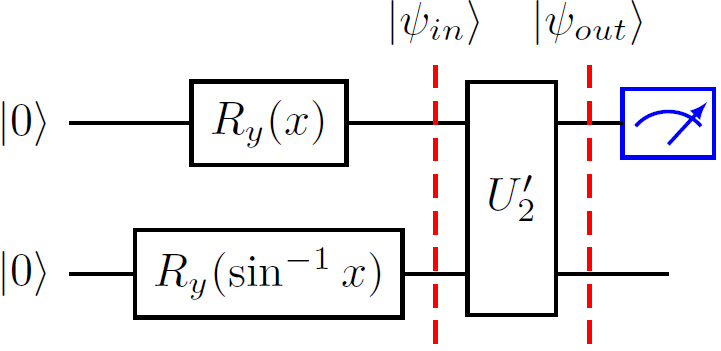}
    \caption{2-qubit, hybrid embedding}
    \label{figure_sub_2qubit_circuit_using_hybrid_embedding}
    \end{subfigure}
\caption{Quantum circuits using different embeddings.}
\label{figure_3_circuits_using_different_embedding}
\end{figure}

\paragraph{Sinusoidal-friendly Embedding.} We propose to directly use $x$ as the input to an angle embedding, which leads to basis functions such as $\sin(x)$ and $\cos(x)$ as detailed below. 

Here we use a \textbf{one-qubit circuit}, shown in Figure \ref{figure_3_circuits_using_different_embedding}a, as a starting point to discuss embedding and measurement. First, rotation operator $R_y$ embeds $x$ into the quantum circuit to generate the quantum input state, $|\psi_{in} \rangle$, as given in Eq. (\ref{equ_input_state}):
\begin{equation}
\label{equ_input_state}
    \begin{split}
        | \psi_{in} \rangle &= R_y (x) |0 \rangle \\
        &= 
        \left[\begin{array}{cc}
          \cos \frac{x}{2} & - \sin \frac{x}{2}\\
          \sin \frac{x}{2} & \cos \frac{x}{2}
        \end{array}\right] 
       \left[\begin{array}{c}
         1\\
         0
       \end{array}\right]\\   
     &= \left[\begin{array}{c}
         \cos \frac{x}{2}\\
         \sin \frac{x}{2}
       \end{array}\right]
    \end{split}
\end{equation}
The motivation to propose this Sinusoidal-friendly embedding is that (1) the Arcsin embedding is not efficient in approximating sinusoidal functions and (2) sinusoidal functions are used in many engineering fields, e.g., they are the heart of AC electrical power systems. 

As shown in Figure \ref{figure_3_circuits_using_different_embedding}a, we use a $2\times 2$ unitary matrix $U$ to represent a one-qubit parameterized quantum circuit, which includes one or more gates. Eq. (\ref{equ_output_state}) provides the vector form of the output state $| \psi_{out} \rangle$ for Figure \ref{figure_3_circuits_using_different_embedding}a. We then use a Pauli Z to measure the output state and the measurement expectation is given in Eq. (\ref{equ_measure}). 
\begin{equation}
\label{equ_output_state}
    \begin{split}
        | \psi_{out} \rangle& = U| \psi_{in} \rangle\\
        &= \left[
        \begin{array}{cc}
             u_{11} & u_{12}\\
             u_{21} & u_{22}
        \end{array}\right] \left[
        \begin{array}{c}
             \cos \frac{x}{2}\\
             \sin \frac{x}{2}
       \end{array}\right]\\
   & = \left[\begin{array}{c}
             u_{11} \cos \frac{x}{2} + u_{12} \sin \frac{x}{2}\\
             u_{21} \cos \frac{x}{2} + u_{22} \sin \frac{x}{2}
       \end{array}\right]
    \end{split}
\end{equation}
\begin{equation}
\label{equ_measure}
        \langle Z \rangle=\langle \psi_{out} |Z| \psi_{out}\rangle=c_2\sin x+c_1\cos x+c_0
\end{equation}
where 
\begin{equation}
\label{equ_c0_c1_c2}
    \begin{split}
     c_0&=(u_{11}^{\ast}u_{11}-u_{21}^{\ast} u_{21}+u_{12}^{\ast}u_{12}-u_{22}^{\ast}u_{22})/2\\
     c_1&=(u_{11}^{\ast}u_{11}-u_{21}^{\ast}u_{21}-u_{12}^{\ast}u_{12}+u_{22}^{\ast}u_{22})/2\\
     c_2&=(u_{11}^{\ast} u_{12} + u_{11} u_{12}^{\ast} -u_{21}^{\ast} u_{22} - u_{21} u_{22}^{\ast})/2
    \end{split}
\end{equation}
where $Z$ is a Pauli-Z matrix (shown in Appendix \ref{section_appen_measure_1st_qubit_of_2qubit_circuit}). 
Eq. (\ref{equ_measure}) shows that the measurement expectation is a linear combination of the three basis functions in $\mathcal{BF}_S=\{1, \sin x, \cos x\}$. 

Similarly, we can calculate the measurement at the first qubit of a \textbf{2-qubit circuit} (depicted in Figure \ref{figure_3_circuits_using_different_embedding}b, the first qubit is on the top). Using Pauli $Z$ to measure the first qubit, we obtain $\langle Z_0\rangle_2$ where subscript $0$ represents a measurement at the first qubit:
\begin{eqnarray}
  \langle Z_0\rangle_2&=&\langle \psi_{out}|Z \otimes I|\psi_{out} \rangle_2  \nonumber\\
  &=&\begin{bmatrix}
      \langle \mathbf{u}_1|\psi_{in}\rangle_2\\
      \langle \mathbf{u}_2|\psi_{in}\rangle_2\\
      \langle \mathbf{u}_3|\psi_{in}\rangle_2\\
      \langle \mathbf{u}_4|\psi_{in}\rangle_2
  \end{bmatrix}^{\dagger}
  \begin{bmatrix}
    1 & 0 &  0 & 0\\
    0 & 1 &  0 & 0\\
    0 & 0 & -1 & 0\\
    0 & 0 &  0 & - 1
  \end{bmatrix}
  \begin{bmatrix}
     \langle \mathbf{u}_1|\psi_{in}\rangle_2\\
     \langle \mathbf{u}_2|\psi_{in}\rangle_2\\
     \langle \mathbf{u}_3|\psi_{in}\rangle_2\\
     \langle \mathbf{u}_4|\psi_{in}\rangle_2
  \end{bmatrix}  \nonumber \\
 &=&\langle\mathbf{u}_1|\psi_{in}\rangle_2^* \langle\mathbf{u}_1|\psi_{in}\rangle_2
   +\langle\mathbf{u}_2|\psi_{in}\rangle_2^* \langle\mathbf{u}_2|\psi_{in}\rangle_2- \nonumber \\
 &&\langle\mathbf{u}_3|\psi_{in}\rangle_2^* \langle\mathbf{u}_3|\psi_{in}\rangle_2
   -\langle\mathbf{u}_4|\psi_{in}\rangle_2^* \langle\mathbf{u}_4|\psi_{in}\rangle_2  \label{equ_Z02_expansion}
\end{eqnarray}
where the subscript 2 indicates a 2-qubit circuit, and the superscripts $\dagger$ and $*$ denote the conjugate transpose and complex conjugate, respectively, $\mathbf{u}_i$ is the $i$th row of matrix $U_2$ with $i=1,2,3,4$, $U_2$ is a $4\times 4$ unitary matrix representing the quantum circuit shown in Figure \ref{figure_3_circuits_using_different_embedding}b. 
Appendix \ref{section_appen_measure_1st_qubit_of_2qubit_circuit} provides detailed calculation of $\langle Z_0\rangle_2$ and shows that using Pauli-Z to measure the first qubit of the 2-qubit circuit (given in Figure \ref{figure_3_circuits_using_different_embedding}b) obtains \textit{9 different basis functions}, i.e., $\mathcal{BF}_2$.
\begin{equation}
\label{equ_BF2}
    \begin{split}
        \mathcal{BF}_2 &= \{1, \sin x_1, \sin x_2, \cos x_1, \cos x_2, \cos x_1\sin x_2,\\
        &\quad\quad\sin x_1\cos x_2, \cos x_1\cos x_2, \sin x_1\sin x_2\}
    \end{split}
\end{equation}

\paragraph{Hybrid Embedding.} To enable our QNN to approximate a general function with both sinusoidal and non-sinusoidal terms, we propose a hybrid embedding. Specifically, we use Arcsin embedding at one qubit and use Sinusoidal-friendly embedding at another qubit. This way, we leverage both Arcsin embedding's strong expressibility and our Sinusoidal-friendly embedding's superiority in expressing sinusoidal-related nonlinear relationships. The hybrid embedding also enables our QNN to be more powerful in ML practice.

Now we analyze the measurement of a \textbf{2-qubit circuit} using the hybrid embedding, as shown in Figure \ref{figure_sub_2qubit_circuit_using_hybrid_embedding}, which embeds $x$ into $R_y(x)$ on the first qubit and $x$ into $R_y(\arcsin x)$ on the second qubit. Then the quantum output state is $|\psi_{out}(x)\rangle=U_2'|\psi_{in}(x)\rangle$ and the expectation of the observable $Z\otimes I$ is $\langle Z\otimes I\rangle=\langle\psi_{in}(x)|(U_2')^{\dagger}(Z\otimes I)U_2'|\psi_{in}(x)\rangle$
where $U_2'$ is the quantum circuit and $|\psi_{in}(x)\rangle=R_y^1(x)\otimes R_y^2(\sin^{-1}x)|00\rangle$. For simplicity of calculation, let $P=(U_2')^{\dagger}(Z\otimes I)U_2'$ and assume that it can be decomposed as:
\begin{equation}
\label{equ_P_equal_UV}
    P=U\otimes V =
    \left[
    \begin{array}{cc}
         u_{11} & u_{12}\\
         u_{21} & u_{22}
    \end{array}\right]  \otimes 
    \left[
    \begin{array}{cc}
         v_{11} & v_{12}\\
         v_{21} & v_{22}
    \end{array}\right]
\end{equation}
Then, Eq. (\ref{equation_mix_embedding}) provides the measurement expectation of the circuit shown in Figure \ref{figure_sub_2qubit_circuit_using_hybrid_embedding}: 
\begin{equation}
\label{equation_mix_embedding}
\begin{split}
    &\quad(R_y^1(x)\otimes R_y^2(\sin^{-1}x)|00\rangle)^{\dagger}P R_y^1(x)\otimes R_y^2(\sin^{-1}x)|00\rangle\\&=(R_y^1(x)|0\rangle)^{\dagger}U(R_y^1(x)|0\rangle)\cdot(R_y^2(\sin^{-1}x)|0\rangle)^{\dagger}V(R_y^2(\sin^{-1}x)|0\rangle)\\
    &=(c_2^\prime\sin x+c_1^\prime\cos x+c_0^\prime)  (d_2x+d_1\sqrt{1-x^2}+d_0)
\end{split}
\end{equation}
where the superscripts 1 and 2 represent qubits 1 and 2, respectively, and 
\begin{equation*}
    \begin{split}
        c_0^\prime&=\frac{u_{11}+u_{22}}{2},\quad c_1^\prime=\frac{u_{11}-u_{22}}{2},\quad c_2^\prime=\frac{u_{12}+u_{21}}{2},\\
        d_0&       =\frac{v_{11}+v_{22}}{2},\quad d_1=\frac{v_{11}-v_{22}}{2},       \quad d_2=\frac{v_{12}+v_{21}}{2}.\\
    \end{split}
\end{equation*}
Eq. (\ref{equation_mix_embedding}) indicates that the quantum circuit using the hybrid embedding, given in Figure \ref{figure_sub_2qubit_circuit_using_hybrid_embedding}, has 9 basis functions, i.e., 
\begin{align}
    \mathcal{BF}_H&=\left\{1, \sin x, \cos x, x, \sqrt{1-x^2}, x \sin x,\sin x \sqrt{1-x^2}, x \cos x,\cos x \sqrt{1-x^2} \right\}
\end{align}
Comparing $\mathcal{BF}_H$ with $\mathcal{BF}_S$ (from the Sinusoidal-friendly embedding) and $\mathcal{BF}_A$ (from the Arcsin embedding) clearly shows that $\mathcal{BF}_A$ and $\mathcal{BF}_S$ are subsets of $\mathcal{BF}_H$ and that basis functions such as $$x \sin x, \sin x \sqrt{1-x^2}, x \cos x, \cos x \sqrt{1-x^2}$$
are unique in $\mathcal{BF}_H$. In summary, \textit{using hybrid embedding can enhance QNNs' expressibility by introducing more different basis functions}.

\subsection{Efficiency of Qubit Usage in QCL}
Assume measuring the output state $|\psi_{out}(x)\rangle=Q(\theta)|\psi_{in}(x)\rangle=Q(\theta)R^Y(\sin^{-1}x)$$|0\rangle$ by an observable $O$, where $Q(\theta)$ represents the quantum circuit. Let $U=Q(\theta)^{\dagger}OQ(\theta)$. Then the expectation of $O$ is
\begin{align}
  \langle O \rangle &= \langle \psi_{out} (x) |O| \psi_{out}
  (x) \rangle\nonumber\\
&=\langle\psi_{in}(x)|Q(\theta)^{\dagger}OQ(\theta)|\psi_{in}(x)\rangle\nonumber\\
  &= (R^Y (\sin^{- 1} x) |0 \rangle)^{\dagger} U R^Y (\sin^{- 1} x) |0
  \rangle\nonumber\\
  &= \left[\begin{array}{cc}
    \cos \left( \frac{\sin^{- 1} x}{2} \right) & \sin \left( \frac{\sin^{- 1}
    x}{2} \right)
  \end{array}\right] \left[\begin{array}{cc}
    u_{11} & u_{12}\\
    u_{21} & u_{22}
  \end{array}\right]\left[\begin{array}{c}
    \cos \left( \frac{\sin^{- 1} x}{2} \right)\\
    \sin \left( \frac{\sin^{- 1} x}{2} \right)
  \end{array}\right]\nonumber\\
  &= u_{11} \cos^2 \left( \frac{\sin^{- 1} x}{2} \right)+u_{22} \sin^2
  \left( \frac{\sin^{- 1} x}{2}\right)\nonumber\\
  &\quad+(u_{12}+u_{21})\cos\left(
  \frac{\sin^{- 1} x}{2} \right) \sin \left( \frac{\sin^{- 1} x}{2} \right)\nonumber\\
  &= \frac{u_{12} + u_{21}}{2} x + \frac{u_{11} - u_{22}}{2} \sqrt{1 -
  x^2} + \frac{u_{11} + u_{22}}{2}
\end{align}
Then applying the linear transformation $w_1\langle O\rangle + w_2$ yields
\begin{align}ax+b\sqrt{1 -x^2}+c\end{align}
where \begin{align*}
    a&=w_1(u_{12}+u_{21})/2\\
    b&=w_1(u_{11}-u_{22})/2\\
    c&=w_1(w_{11}+w_{22})/2+w_2
\end{align*}
The above analysis clearly shows that the quantum circuits utilizing one qubit can only express functions consisting of linear combinations of $1,x,\sqrt{1-x^2}$. Next, we theoretically analyze why these quantum circuits cannot approximate $\sin x$ well.

\paragraph{Error Analysis.}
Based on Taylor expansion, we have:
\begin{equation}\label{tysinx}
    \sin x = x - \frac{x^3}{3!} + \frac{x^5}{5!} + o (x^5)
\end{equation}
\begin{align} \sqrt{1 - x^2} = 1 - \frac{1}{2} x^2 - \frac{1}{4} x^4 + o (x^4)\end{align}
Then the linear transformation of the measurement expectation is,
\begin{equation}\label{tysqrt}
    ax+b\sqrt{1 -x^2}+c=b+c+ax-\frac{b}{2}x^2-\frac{b}{4}x^4+o(x^4)
\end{equation}
To minimize the error of using the right-hand side of Eq. (\ref{tysqrt}) to
approximate Eq. (\ref{tysinx}), we have 
\begin{align}b + c = 0, a = 1 \Rightarrow c = - b, a = 1\end{align}
as $x\in(-1,1)$. 
Then we have:
\begin{align}
  \sin x-(ax+b\sqrt{1 -x^2}+c)&=\sin x-(x +b \sqrt{1 - x^2} - b)\nonumber\\
  & = x -\frac{x^3}{3!} + \frac{x^5}{5!} 
  + o(x^5)-\left[ x-\frac{b}{2} x^2 -\frac{b}{4} x^4 + o (x^4) \right]\nonumber\\
  & =  \frac{b}{2} x^2 + \frac{b}{4} x^4 + o (x^4)  - \frac{x^3}{3!} +
  \frac{x^5}{5!} + o(x^5)\nonumber\\
  & =  \left( \frac{b}{2} - \frac{x}{3!} \right) x^2 + \left( \frac{b}{4} -
  \frac{x}{5!} \right)x^4 + o(x^4)\nonumber\\
  & = O(x^2)
\end{align}

That is, the error of the best function expressed by QCL utilizing one qubit to approximate $\sin x$ is in the scale of $O(x^2)$ which is non-negligible in practice.

\begin{figure}
     \centering
     \begin{subfigure}[b]{0.45\textwidth}
         \centering
         \includegraphics[width=\textwidth]{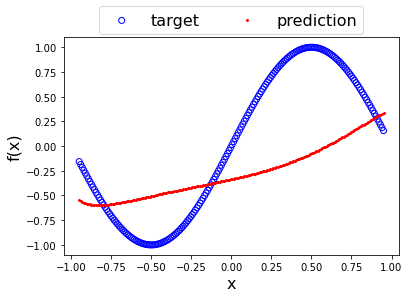}
         \caption{one qubit.}
         \label{qcl_one_qubit_f1v1}
     \end{subfigure}
      \begin{subfigure}[b]{0.45\textwidth}
         \centering
         \includegraphics[width=\textwidth]{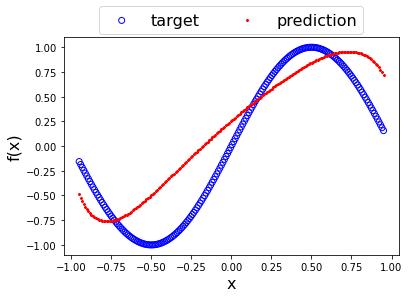}
         \caption{two qubits.}
         \label{qcl_two_qubit_f1v1}
     \end{subfigure}
     \caption{Fitting result using QCL}
\end{figure}

Figure \ref{qcl_one_qubit_f1v1} shows the result of one-qubit QCL in fitting $\sin(\pi x)$, which indicates QCL completely fails to learn $\sin(\pi x)$. Probably by changing the learning rate, the number of epochs, or other hyper-parameters we could get a slightly better result, but the QCL using one qubit still cannot succeed in fitting $\sin(\pi x)$.

We can similarly analyze the approximation error of QCL using two qubits, but not given here to save space. 
Figure \ref{qcl_two_qubit_f1v1} shows the result of two-qubit QCL in fitting $\sin(\pi x)$, which demonstrates that QCL using two qubits still cannot approximate $\sin(\pi x)$ effectively.

From theoretical and experimental analysis, we conclude that \textbf{Arcsin Embedding} is not a qubit-efficient approach to express sinusoid functions. In high-dimensional space, this issue is severer. This motivated us to propose the sinusoidal-friendly embedding and the hybrid embedding described above.

\subsection{Strategy 2: Redundant Measurement}
\label{method_strategy2_redundant_measurement}
This section investigates the impact of redundant measurements on QNNs' performance theoretically and experimentally. 

We tend to fall into the trap: redundant measurements could benefit the performance of QNNs as extra measurement could extract more information from the quantum output state. But generally, this is not true. In fact, as the analysis in the following paragraph shows, measuring two quantum bits by Pauli Z produces the same basis functions as measuring one quantum bit. That is, theoretically, the information extracted from the quantum output state is identical. Therefore, two QNNs should have the same power to approximate functions. But the experimental results in Table \ref{table_ablation_study_all_3_method} show that there could be an obvious difference in QNNs' performance to fit a given function using one and two qubits, which breaks our intuition. 

Let's start the analysis with a simple example. If we use Pauli Z to measure the second qubit of the 2-qubit circuit given in Figure \ref{figure_3_circuits_using_different_embedding}b, similar to the analysis given in Section \ref{section_appen_measure_1st_qubit_of_2qubit_circuit}, we can obtain the measurement expectation as $\langle Z_1\rangle_2$ where subscript 1 represents a measurement at the second qubit:
\begin{eqnarray}
\langle Z_1\rangle_2 &=& \langle \psi_{out} | I \otimes Z|\psi_{out} \rangle_2  \nonumber \\
  &=&\begin{bmatrix}
      \langle \mathbf{u}_1|\psi_{in}\rangle_2\\
      \langle \mathbf{u}_2|\psi_{in}\rangle_2\\
      \langle \mathbf{u}_3|\psi_{in}\rangle_2\\
      \langle \mathbf{u}_4|\psi_{in}\rangle_2\end{bmatrix}^{\dagger}
  \begin{bmatrix}
    1 &  0 & 0 & 0\\
    0 & -1 & 0 & 0\\
    0 &  0 & 1 & 0\\
    0 &  0 & 0 & -1
  \end{bmatrix}
 \begin{bmatrix}
     \langle \mathbf{u}_1|\psi_{in}\rangle_2\\
     \langle \mathbf{u}_2|\psi_{in}\rangle_2\\
     \langle \mathbf{u}_3|\psi_{in}\rangle_2\\
     \langle \mathbf{u}_4|\psi_{in}\rangle_2\end{bmatrix}  \nonumber \\
 &=& \langle\mathbf{u}_1|\psi_{in}\rangle_2 ^*\langle\mathbf{u}_1|\psi_{in}\rangle_2
     -\langle\mathbf{u}_2|\psi_{in}\rangle_2 ^*\langle\mathbf{u}_2|\psi_{in}\rangle_2  \nonumber \\
 && +\langle\mathbf{u}_3|\psi_{in}\rangle_2 ^*\langle\mathbf{u}_3|\psi_{in}\rangle_2
     -\langle\mathbf{u}_4|\psi_{in}\rangle_2 ^*\langle\mathbf{u}_4|\psi_{in}\rangle_2   \label{equation_Z1_measure_2nd_qubit}
\end{eqnarray}
Comparing Eqs. (\ref{equation_Z1_measure_2nd_qubit}) with (\ref{equ_Z02_expansion}), we see that their only difference is the coefficients of the second and the third terms. 
Obviously, these two equations have the same basis functions which are shown in Eq. (\ref{equ_BF2}). 
That is, measuring the second qubit yields the same basis functions as measuring the first qubit. 
Therefore, \textit{even if we measure both qubits, we cannot obtain more basis functions than measuring only one qubit}. This is generally valid if the observable is a diagonal matrix. 

As our \textbf{redundant measurement strategy}, we use a weighted sum of both measurements, $\langle Z_{ws} \rangle_2$, to represent the output of the 2-qubit quantum circuit (given in Figure \ref{figure_3_circuits_using_different_embedding}b), which is described in Eq. (\ref{equ_weighted_sum_of_two_measurements}):
\begin{equation}
\label{equ_weighted_sum_of_two_measurements}
    \langle Z_{ws} \rangle_2 = w_0\langle Z_0\rangle_2 + w_1\langle Z_1\rangle_2+w_2
\end{equation}
where $w_0, w_1$ and $w_2$ are three optimizing variables.

Since $\langle Z_0 \rangle_2$ and $\langle Z_1 \rangle_2$ have the same basis functions, being a weighted sum of them as shown in Eq. (\ref{equ_weighted_sum_of_two_measurements}), $\langle Z_{ws} \rangle_2$ also have the same basis functions. However, $\langle Z_{ws} \rangle_2$ introduces one more parameter, i.e., $w_3$ in Eq. (\ref{equ_weighted_sum_of_two_measurements}), into QNNs. 
Considering that either $\langle Z_0 \rangle_2$ or $\langle Z_1 \rangle_2$ can be expressed as a weighted sum of basis functions, $\langle Z_{ws} \rangle_2$ is also a weighted sum of the same basis functions but has one more parameter, $w_1$. In other words, using the redundant measurement introduces one more parameter into the coefficients of basis functions. This new parameter will be optimized in each epoch during training QNNs. This creates more flexibility in QNNs and makes them more expressive, which has been verified in our experiments. 

In summary, \textit{if the observable is a diagonal matrix, using the redundant measurement does not increase the number of different basis functions, but enhances QNNs' expressibility via increasing the coefficient flexibility of basis functions.}

\textbf{Remark 1. }As detailed in Appendix \ref{section_appen_number_of_basis_functions}, \textit{by measuring one qubit of its $n$-qubit circuit, a QNN can have $3^n$ different basis functions. That is, due to the tensor product operation, \textbf{the number of different basis functions grows exponentially as the number of qubits increases}}, which implies that \textit{QNN can have an exponential advantage over classical NNs in terms of expressibility}.

\textbf{Remark 2.} We want to emphasize that \textbf{even though more measurements do not always increase the number of base functions and further enhance QNNs' theoretical capacity to express functions, in practice it is still beneficial to measure multiple qubits as resources allow}. For example, our QNN can approximate $\sin(\pi x)$ with zero error (theoretically) by one measurement as seen in Eq. (\ref{equ_post_measure_Z_Zsquare}). But measuring two qubits separately brings a significant improvement compared to measuring one qubit, as shown in Table \ref{table_ablation_study_all_3_method} (the comparison between QNN-exc2 and QNN-A, e.g., the 2nd row vs. the 5th row). It would be interesting to further study skills of designing observable matrices so that redundant measurements can generate more basis functions. 

\subsection{Strategy 3: Post-Measurement Function}
\label{method_strategy3_post_measurement_function}

Note that the measurement from a 1-qubit circuit, as shown in Eq. (\ref{equ_measure}), does not generate second-order basis functions such as $\sin^2 x, \cos^2 x, \sin x \cos x$.  To enable 1-qubit QNNs to have these basis functions, we propose to use a \textbf{post-measurement function}, $T_w(\langle Z \rangle)$, instead of using the measurement $\langle Z \rangle$ itself as the Output $f_\theta(x)$, as depicted in Figure \ref{figure_QNN_diagram}. For example, we set $T_w(\langle Z \rangle)$ as a second-order polynomial function of $\langle Z \rangle$, as shown in Eq. (\ref{equ_post_measure_Z_Zsquare}), where Eq. (\ref{equ_measure}) is used to expand the $\langle Z \rangle$:
\begin{equation}
\label{equ_post_measure_Z_Zsquare}
    \begin{split}
    T_w (\langle Z \rangle) &= w_0 + w_1\langle Z\rangle+w_2\langle Z\rangle^2\\& = e_0 + e_1\cos x+ e_2\sin x + e_3\sin x\cos x + e_4\cos^2x + e_5\sin^2x
    \end{split}
\end{equation}
where  
\begin{align*}
    e_0&= w_1 (c_0 + 2 w_1 w_2 c_0^2)+w_0 & e_1&=w_1 c_1 (1 + 2 w_1 w_2 c_0)\\
    e_2&= w_1 c_2 (1 + 2 w_1 w_2 c_0) & e_3&=2 w_1^2 w_2 c_2 c_0\\
    e_4&=w_1^2 w_2 c_1^2              & e_5&=w_1^2 w_2 c_2^2 
\end{align*}
where $c_0, c_1$ and $c_2$ are given in Eq. (\ref{equ_c0_c1_c2}). 

Note that the basis functions are $\mathcal{BF}_s$ (given in the paragraph after Eq. (\ref{equ_c0_c1_c2})) when no post-measurement function is used for the circuit given in Figure \ref{figure_3_circuits_using_different_embedding}a. 
Eq. (\ref{equ_post_measure_Z_Zsquare}) indicates that the basis functions expand from $\mathcal{BF}_S$ to $\mathcal{BF}_P=\{1, \cos x, \sin x, \sin x \cos x, \cos^2x, \sin^2x\}$. \textit{In short, using the post-measurement function expands the basis functions}. Note that this function appears in the classical part of QNN ($T_w$ in Figure \ref{figure_QNN_diagram}), i.e., it does not need quantum resources,
and that we need to simultaneously optimize $w_0, w_1$ and $w_2$ when training QNNs, as depicted in Figure \ref{figure_QNN_diagram}.

The post-measurement function can take any form as required. Eq. (\ref{equ_post_measure_Z_Zsquare}) is just one example of the post-measurement function. Another example is adding higher-order terms (such as $\langle Z \rangle^3$ and $\langle Z \rangle^4$) or other nonlinear terms (such as $e^{\langle Z \rangle}, \sqrt{\langle Z \rangle}$, and $|\langle Z \rangle|$) to the post-measurement function to generate more nonlinear basis functions. It is worth noting that problem-specific knowledge can help to design a more appropriate and effective post-measurement function.

Thus, we believe that \textit{using post-measurement functions is an excellent strategy to introduce more nonlinear operations into QNNs to enhance their expressibility without using additional quantum resources such as quantum gates and qubits}. Note that this strategy helps to reduce the required quantum circuit depth/size, and therefore helps to speed up the application of QNNs in the current NISQ era.

\subsection{Strategy 4: Random Training Data}
\label{method_strategy4_random_training_data}
Here, we present two methods to generate training data for QNNs. 
\textbf{To generate Meshgrid data}, we uniformly sample data from the input space. Specifically, the sample data are uniformly spaced in each dimension of the input space. We simply use the `meshgrid' command in Python to generate the Meshgrid data.

\textbf{To generate random training data}, 
we draw samples $\mathbf{x}$ from a uniform distribution and then generate $y$ from a given function $y=f(\mathbf{x})$. We then use the set of pairs $\{(\mathbf{x}, y)\}$ as training data.

\textbf{Remark 3. }Sinusoidal functions are of theoretical importance as they can well approximate many analytical functions, especially periodic functions \cite{Rahimi2008}. In addition, they are widely used in engineering fields. For example, AC power systems, including generation and grid, are based on sinusoidal functions \cite{Glover2011powerSystemBook}. We believe it would be very beneficial to employ Sinusoidal-friendly or hybrid embeddings in QNNs to solve power system ML problems \cite{Xie_2020MLreviewPowerSystem, Miraftabzadeh_2019MLreviewPowerSystem, Zhou2022QuantumComputingInPowerSystem}.

\section{Experiments and Results}
\label{target_function}

In our experiments below, the QNNs with different strategies are compared to validate the effectiveness of each strategy. For the convenience of expression, we define \textit{different variants of QNNs} here: \\
$\bullet$ QNN-A: QNN with all four strategies. \\
$\bullet$ QNN-A2: QNN with all four strategies using the hybrid embedding. \\
$\bullet$ QNN-exc1: QNN with strategies 2, 3, and 4, and using the Arcsin embedding.\\
$\bullet$ QNN-exc2: QNN with strategies 1, 3, and 4, i.e., no redundant measurement.\\
$\bullet$ QNN-exc3: QNN with strategies 1, 2, and 4, i.e., no post-measurement function.\\
$\bullet$ QNN-exc4: QNN with strategies 1, 2, and 3, i.e., use mesh grid data instead of using random training data.\\
$\bullet$ QNN-exc5: QNN with only strategy 1.

Note that QNN-A, QNN-exc2, QNN-exc3, QNN-exc4, and QNN-exc5 all use Sinusoidal-friendly embedding.

In QNN variants, we use Adam \cite{Kingma2014} as the optimizer because it usually performs well in different ML tasks. We initialize QNN parameters using uniformly random data between 0 and $2\pi$. We set the initial value of each of $w_0, w_1$, and $w_2$ to 0. We set the input space $\mathcal{X}=[-0.95,0.95]^{d}$ such that we can compare our  method with QCL \cite{mitarai2018QCL}, where $d$ is the dimension of input variables. All errors given below represent the mean absolute error \footnote{\url{https://en.wikipedia.org/wiki/Mean_absolute_error}} between the predicted and the true values.

\subsection{Experiment Design}
To numerically validate the effectiveness of each strategy, we compare the performance of different variants of QNNS and QCL in approximating different target functions. 

\begin{figure*}[t]
\centering
    \begin{subfigure}[b]{0.318\textwidth}
    \centering
    \includegraphics[width=\textwidth]{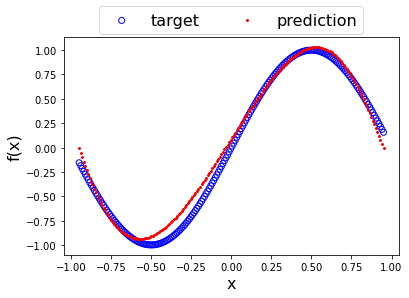}
    \caption{QCL}
    \end{subfigure} 
    \begin{subfigure}[b]{0.318\textwidth}
    \centering
    \includegraphics[width=\textwidth]{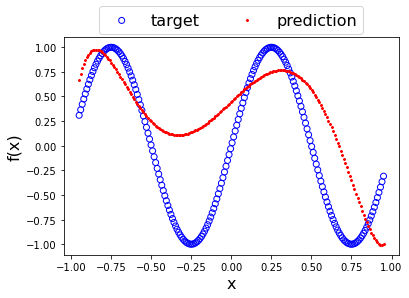}
    \caption{QCL}
    \end{subfigure} 
    \begin{subfigure}[b]{0.318\textwidth}
    \centering
    \includegraphics[width=\textwidth]{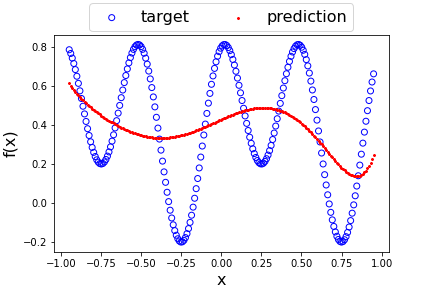}
    \caption{QCL}
    \end{subfigure} 
    \begin{subfigure}[b]{0.318\textwidth}
    \centering
    \includegraphics[width=\textwidth]{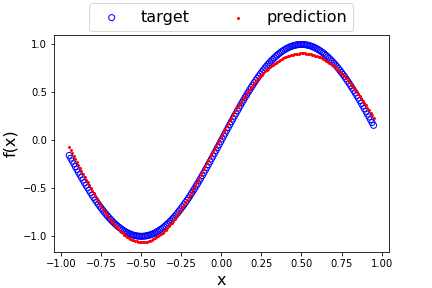}
    \caption{QNN-exc2}
    \end{subfigure} 
    \begin{subfigure}[b]{0.318\textwidth}
    \centering
    \includegraphics[width=\textwidth]{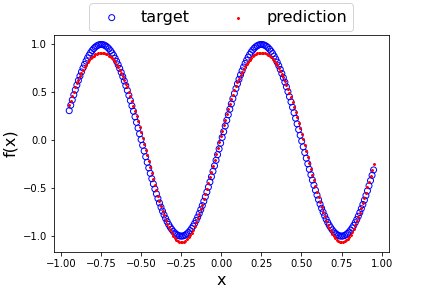}
    \caption{QNN-exc2}
    \end{subfigure} 
    \begin{subfigure}[b]{0.318\textwidth}
    \centering
    \includegraphics[width=\textwidth]{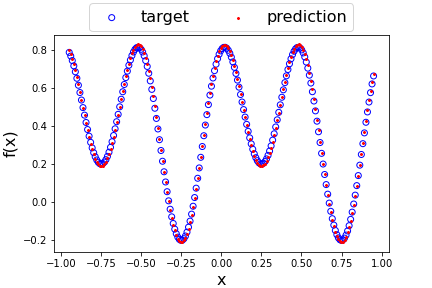}
    \caption{QNN-exc2}
    \end{subfigure}
\caption{Fitting results of two methods on different functions: left panel (a and d) use $f_{1v1}$,  middle panel (b and e) use $f_{1v2}$,  right panel (c and f) use $f_{1v3}$, where 'target' represents true values, 'prediction' represents predicted values.}
\label{figure_fitting_result_QCL_vs_QNN}
\end{figure*}

We list all the functions used in this paper in Eqs. (\ref{equ_obj_f1v1})-(\ref{equ_obj_f3}). For the convenience of expression, we drop the `$(x)$' and the `$(\mathbf{x})$' when referring to these 6 target functions in the rest of the paper without introducing ambiguity. Note that the function name's first number in the subscript denotes the dimension of the function variable, e.g., $f_{1v2}, f_2$ and $f_3$ are 1-, 2-, and 3-variable functions, respectively. 
\begin{align}
    f_{1v1}(x)&=\sin(\pi x)\label{equ_obj_f1v1}\\
    f_{1v2}(x)&=\sin(2\pi x)\label{equ_obj_f1v2}\\
    f_{1v3}(x)&=0.2\sin(2\pi x)+0.8\cos^2(2\pi x)\label{equ_obj_f1v3}\\
    f_{1v0}(x)&=\sin(2\pi x)+0.5\sqrt{1-x^2}+x\label{equ_obj_f1v0}\\
    f_2(\mathbf{x})&=0.5\sin(\pi x_1)\sin(\pi x_2))+\quad0.8\cos^2(\pi x_1) + 0.3\sin(\pi x_2)\label{equ_obj_f2}\\
    f_3(\mathbf{x})&=0.5\sin(x_1) \sin(x_2) \quad0.6\cos(x_2)\sin(x_3)+\cos^2(x_3)\label{equ_obj_f3}
\end{align}

\subsection{Comparison Between QNN and QCL}
\label{section_comparison_between_QNN_and_QCL}

\begin{figure*}[t]  
\centering
    \begin{subfigure}[b]{0.301\textwidth}
    \centering
    \includegraphics[width=\textwidth]{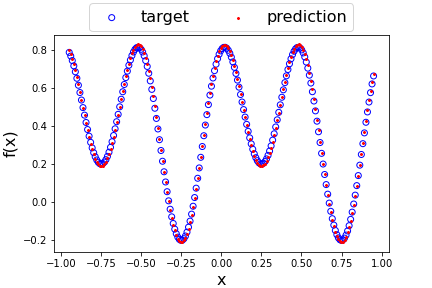}
    \caption{QNN-exc2} 
    \end{subfigure}
    \begin{subfigure}[b]{0.331\textwidth}
    \centering
    \includegraphics[width=\textwidth]{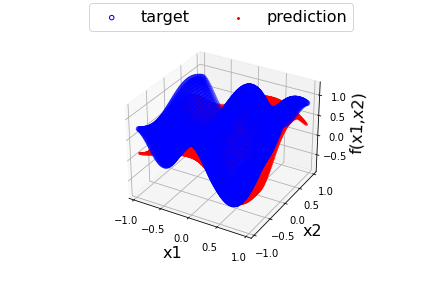}
    \caption{QNN-exc2} 
    \end{subfigure}
    \begin{subfigure}[b]{0.301\textwidth}
    \centering
    \includegraphics[width=\textwidth]{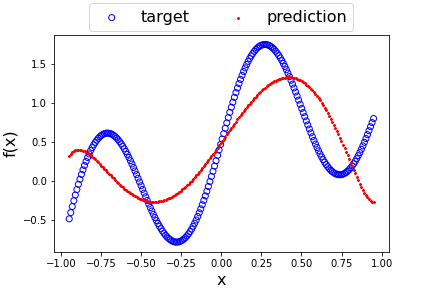}
    \caption{QNN-exc1}
    \end{subfigure}
\newline
    \begin{subfigure}[b]{0.301\textwidth}
    \centering
    \includegraphics[width=\textwidth]{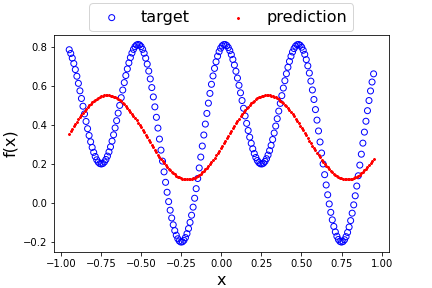} 
    \caption{QNN-exc3} 
    \end{subfigure}
    \begin{subfigure}[b]{0.331\textwidth}
    \centering
    \includegraphics[width=\textwidth]{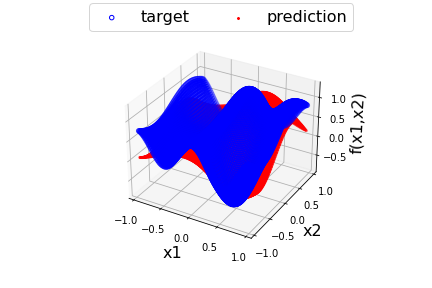}
    \caption{QNN-exc3} 
    \end{subfigure}
    \begin{subfigure}[b]{0.301\textwidth}
    \centering
    \includegraphics[width=\textwidth]{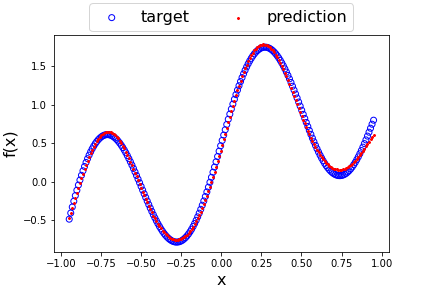}
    \caption{QNN-A2}
    \end{subfigure}
\newline
    \begin{subfigure}[b]{0.301\textwidth}
    \centering
    \includegraphics[width=\textwidth]{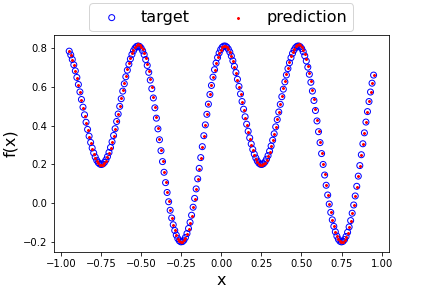} 
    \caption{QNN-exc4} 
    \end{subfigure}
    \begin{subfigure}[b]{0.331\textwidth}
    \centering
    \includegraphics[width=\textwidth]{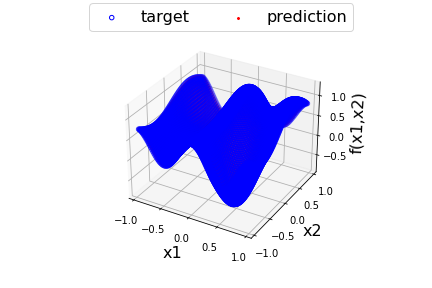}
    \caption{QNN-exc4 }
    \end{subfigure}
    \begin{subfigure}[b]{0.28\textwidth}
    \centering
    \includegraphics[width=\textwidth]{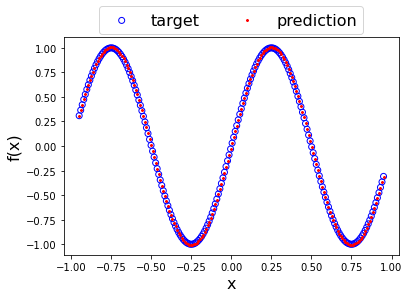} 
    \caption{QNN-exc5}
    \end{subfigure}
\newline
    \begin{subfigure}[b]{0.301\textwidth}
    \centering
    \includegraphics[width=\textwidth]{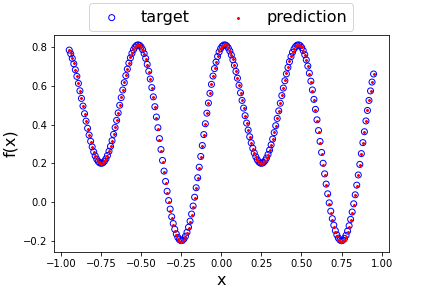} 
    \caption{QNN-A} 
    \end{subfigure}
    \begin{subfigure}[b]{0.331\textwidth}
    \centering
    \includegraphics[width=\textwidth]{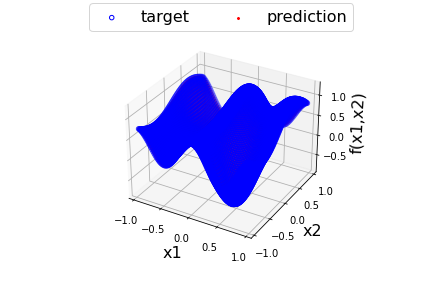}
    \caption{QNN-A} 
    \end{subfigure}
\begin{subfigure}[b]{0.331\textwidth}
    \centering
    \includegraphics[width=\textwidth]{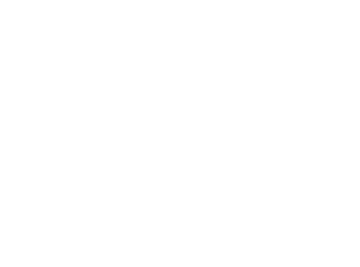}
    \end{subfigure}
\caption{Fitting results from different QNN variants on different functions: left panel (a, d, g, j) use function $f_{1v3}$,  middle panel (b, e, h, k) use function $f_2$, subfigures c and f use function $f_{1v0}$, and subfigure i uses the function $f_{1v2}$.}
\label{figure_fitting_result_11sub_figures}
\end{figure*}

We use three functions with different complexity to validate the performance of QCL \cite{mitarai2018QCL} and our QNN-exc2: $f_{1v1}$ is the simplest. $f_{1v2}$ is more complex as it has two cycles. $f_{1v3}$ is more complex than the previous two as it has a second-order term. We use QNN-exc2 instead of QNN-A because QCL also uses only one measurement.

Figures \ref{figure_fitting_result_QCL_vs_QNN}a and \ref{figure_fitting_result_QCL_vs_QNN}d show that QNN-exc2 achieves slightly better performance than QCL\footnote{The code for QCL is downloaded from \url{https://github.com/dawidkopczyk/qcl}. Paper \cite{mitarai2018QCL} writes the target function as $\sin x$ but should be sin($\pi x$) based on the figure shown.} on function $f_{1v1}$. Moreover, QNN-exc2 utilizes only $2$ qubits while QCL uses $3$. 
Figures \ref{figure_fitting_result_QCL_vs_QNN}b and \ref{figure_fitting_result_QCL_vs_QNN}e indicate that QNNs-exc2's performance in learning $f_{1v2}$ is much better than that of QCL. It also reveals that the QCL cannot fit $\sin(2\pi x)$ which is a simple univariate function. 
Figures \ref{figure_fitting_result_QCL_vs_QNN}c and \ref{figure_fitting_result_QCL_vs_QNN}f clearly reveal that QCL cannot fit function $f_{1v3}$ while our method works perfectly. In summary, Figure \ref{figure_fitting_result_QCL_vs_QNN} shows that \textit{our QNN outperforms QCL, and the superiority becomes more obvious as the function complexity increases}. 

\subsection{Impact of Strategy 1 - Embedding}

\textbf{Validate Sinusoidal-friendly embedding. }
For a fair comparison, we compare QCL with QNN-exc5 (it excludes strategies 2, 3, and 4) in approximating function $f_{1v2}$. Their main difference is that the former uses Arcsin embedding, while the latter uses Sinusoidal-friendly embedding. 

QNN-exc5 uses a simple quantum circuit with 2 qubits and 4 layers, as shown in Figure \ref{figure_appendix_quantum_circuit_f1v3}. 
We show QCL and QNN-exc5's fitting results in Figures \ref{figure_fitting_result_QCL_vs_QNN}b and \ref{figure_fitting_result_11sub_figures}i, respectively. The two figures clearly indicate the superior performance of QNN-exc5 over QCL, which \textit{validates the superiority of Sinusoidal-friendly embedding over the Arcsin embedding} in fitting $f_{1v2}$, a sinusoidal-based function.

\paragraph{Validate hybrid embedding.} We compare QNN-A2 and QNN-exc1's performance on approximating function $f_{1v0}$ which has a sinusoidal term and non-sinusoidal terms such as $\sqrt{1-x^2}$. 
Each QNN variant uses 2 qubits. In QNN-A2, one qubit uses the Sinusoidal-friendly embedding and the other uses the Arcsin embedding. In QNN-exc1, both qubits use Arcsin embedding.

Figure \ref{figure_fitting_result_11sub_figures}f demonstrates QNN-A2's excellent performance in fitting $f_{1v0}$, which indicates that \textit{\textbf{the hybrid embedding enables our QNN-A2 to fit general functions with both sinusoidal and non-sinusoidal terms}}.

Figure \ref{figure_fitting_result_11sub_figures}c shows that QNN-exc1 does not fit $f_{1v0}$ well. Since the only difference between QNN-A2 and QNN-exc1 is the embedding method at one of the two qubits, these two figures again \textit{confirm the superiority of using Sinusoidal-friendly embedding over Arcsin embedding to express nonlinear relationships with sinusoidal components}.

\subsection{Ablation Study}
To validate the effectiveness of strategies 2$\sim $4, we conduct ablation studies on functions $f_{1v3}, f_2$, and $f_3$. For a fair comparison, different QNN variants use the same quantum circuit on the same function. As higher-dimensional functions are more complex, we use 2, 3, and 4 qubits for the three functions, respectively. All quantum circuits use four layers, as shown in Figures \ref{figure_appendix_quantum_circuit_f1v3} 
and \ref{figure_appendix_quantum_circuit_f3}).
\paragraph{From Table \ref{table_ablation_study_all_3_method}: }
(1) The last column shows that strategy 3 dramatically increases QNN's fitting accuracy on $f_{1v3}$, that strategies 2 and 3 each significantly enhance QNN's performance on fitting $f_2$, and that in other cases, a strategy helps to reduce fitting errors by 1.1$\sim$3.6 times. 
(2) The training and test errors are very close except QNN-exc4 on $f_3$. This shows that different variants of QNNs generally do not have overfitting problems on the three functions. QNN-exc4 has an overfitting problem (i.e., test error being much higher than training error) on $f_3$ while QNN-A does not, which implies using random training data can help to avoid overfitting problems compared to using Meshgrid data.
\paragraph{From Figure \ref{figure_fitting_result_11sub_figures}: }
(1) QNN-A performs much better than QNN-exc3 on fitting function $f_{1v3}$ as shown on the left panel. QNN-A clearly outperforms QNN-exc2 and QNN-exc3 on approximating function $f_2$ as shown on the middle panel. This matches well with the previous paragraph's first observation.
(2) Figures 
\ref{figure_fitting_result_11sub_figures}a, 
\ref{figure_fitting_result_11sub_figures}g, and 
\ref{figure_fitting_result_11sub_figures}j look similar. 
Figures 
\ref{figure_fitting_result_11sub_figures}h and 
\ref{figure_fitting_result_11sub_figures}k also seem to be the same. However, the fitting errors associated with these figures are different as described in the previous paragraph. Note that rows 2$\sim$9 of Table \ref{table_ablation_study_all_3_method} are obtained by the same methods on the same functions as in Figures 
\ref{figure_fitting_result_11sub_figures}a, \ref{figure_fitting_result_11sub_figures}d, \ref{figure_fitting_result_11sub_figures}g,  \ref{figure_fitting_result_11sub_figures}j, \ref{figure_fitting_result_11sub_figures}b, \ref{figure_fitting_result_11sub_figures}e, \ref{figure_fitting_result_11sub_figures}h, and \ref{figure_fitting_result_11sub_figures}k, respectively. 

In conclusion, the results show that each of strategies 2, 3, and 4 can significantly improve our QNN's ability to learn nonlinear functions and that QNN-A outperforms all other three variants in all three functions.
\begin{table}[t]
    \centering
    \begin{tabular}{lcccc}
     \toprule
      Method   & Func.  & Training error        & Test error & rat. \\
      \hline
          QNN-exc2 & $f_{1v3}$ & 8.07$\cdot 10^{-3}$ & 7.99$\cdot 10^{-3}$ & 2.9 \\
          QNN-exc3 & $f_{1v3}$ & 2.17$\cdot 10^{-1}$ & 2.48$\cdot 10^{-1}$ & 92 \\
          QNN-exc4 & $f_{1v3}$ & 3.00$\cdot 10^{-3}$ & 3.01$\cdot 10^{-3}$ & 1.1 \\ 
          QNN-A    & $f_{1v3}$ & 2.58$\cdot 10^{-3}$ & 2.71$\cdot 10^{-3}$ & -\\
      \hline
          QNN-exc2 & $f_2$     & 2.28$\cdot 10^{-1}$ & 2.35$\cdot 10^{-1}$ & 37\\
          QNN-exc3 & $f_2$     & 2.36$\cdot 10^{-1}$ & 2.54$\cdot 10^{-1}$ & 40\\
          QNN-exc4 & $f_2$     & 8.21$\cdot 10^{-3}$ & 8.60$\cdot 10^{-3}$ & 1.3\\
          QNN-A    & $f_2$     & 6.05$\cdot 10^{-3}$ & 6.40$\cdot 10^{-3}$ & -\\
      \hline
          QNN-exc2 & $f_3$     & 1.35$\cdot 10^{-1}$ & 1.40$\cdot 10^{-1}$ & 3.4\\
          QNN-exc3 & $f_3$     & 4.79$\cdot 10^{-2}$ & 4.93$\cdot 10^{-2}$ & 1.2\\
          QNN-exc4 & $f_3$     & 2.71$\cdot 10^{-2}$ & 1.48$\cdot 10^{-1}$ & 3.6\\
          QNN-A    & $f_3$     & 3.75$\cdot 10^{-2}$ & 4.08$\cdot 10^{-2}$ & -\\
    \bottomrule
    \end{tabular}
    \caption{Training/test error of different variants of QNN on three functions, where `rat.' represents the ratio between the test error of the corresponding method and that of QNN-A on the same function. For example, the ratio=92 in the third row of the table is obtained from $(2.48\cdot 10^{-1}) \div (2.71\cdot 10^{-3})$.}
    \label{table_ablation_study_all_3_method}
\end{table}

\subsection{Variance Analysis}
\label{section_variance_analysis}
We analyze the impacts of using random training data on the performance variance of QNN-A. We run QNN-A 150 times independently, without using a seed method to customize the start number of the random number generator. That is, the training data in different runs are independent of each other, but all follow the same uniform random distribution. 

To test this impact, we use QNN-A to approximate two functions, $f_{1v3}$ and $f_{1v0}$. We collect the mean value of the test errors in each run, i.e., we collect one error value from each run. The mean and variance of the 150 values from 150 runs are $0.012$ and $5.19\times 10^{-4}$ for the function $f_{1v3}$ and $0.021$ and $1.18\times 10^{-4}$ for the function $f_{1v0}$, respectively. 
The variance in each case is very small, which validates the QNN-A's stable performance in using random training data. 
The mean values being 0.021 and 0.012 indicate that the average fitting performance of QNN-A is very good. More analysis and results, such as histograms, are given in Appendix \ref{section_appen_variance_analysis}.

\subsection{Application for Real-Word Task}
In this section, we benchmark our proposed strategies on an open-access clinical dataset, i.e., Heart failure clinical records Data Set \cite{chicco2020machine}. 
This dataset contains the medical records of 299 patients, of which each has 13 clinical features, denoted as  $x_1,x_2,\cdots,x_{12}, y$. A machine learning model is designed to predict the occurrence of the death event, represented by $y=1$ or $0$, using the remaining 12 features as input. 
More details of the dataset are given in Appendix \ref{section:dataset_description}.

\paragraph{Data Processing}
Since the input features $x_1,x_2,\cdots,x_{12}$ are not on the same scale, it is necessary to normalize the raw data before feeding it into the QNN model. We propose to use the following normalization formulas: $x_1/100$, $\ln(1+x_3)/8$, $x_5/100$, $x_7/300000$, $x_8/150$, $x_9/2$, and $\ln(1+x_{12})/5$. Also, the target $y$ is transformed by $2y-1$.
\paragraph{Implementation}
After data normalization, we divide the entire dataset into 5 folders for the leave-one-out cross-validation, using the StratifiedKFold algorithm\footnote{\url{https://scikit-learn.sourceforge.net/dev/modules/generated/sklearn.cross_validation.StratifiedKFold.html}}. Due to the imbalanced nature of the dataset (the ratio between death and living instances is $96:203$), we re-sample the minority class such that the two classes have the same number of instances in the training stage. Specifically, we use either the uniformly randomly repeating the minority instances (URRM) or the Synthetic Minority Oversampling Technique (SMOTE) \cite{chawla2002smote}. We use the same setting (such as network structure and optimizer) and strategies as the study on the synthetic datasets, while excluding hybrid embedding and random training data. 
To evaluate our QNN's performance, we use the cross-validation area under the receiver operator characteristics curve (AUC) and classification error metrics. 
To ensure the validity of our approach, we compared our results to those reported in \cite{moradi2022clinical}, with the key difference being the usage of more features and qubits in our model. As shown in Table \ref{model performance}, our QNN-A demonstrates superior performance with low classification error and much higher AUC compared to \cite{moradi2022clinical}. 

\begin{table}[t]
    \centering
    \begin{tabular}{lccc}
     \toprule
      Method  & qubits & AUC  &  Error rate\\
      \hline
      \multirow{2}{*}{qDC} & 3 & 0.62 & NA\\
      \cline{2-4}
       & 8 & 0.60 & NA\\
      \hline
      \multirow{2}{*}{qKSVM}& 3 & 0.51 & NA\\
      \cline{2-4}
      & 8 & 0.51 & NA\\
      \hline
      \multirow{2}{*}{sqKSVM}& 3 & 0.51 & NA\\
      \cline{2-4}
      & 8 & 0.51 & NA\\
      \hline
      \multirow{2}{*}{QNN-A}& 13 & 0.90 & 0.19\\
      \cline{2-4}
      & 13 & 0.85 & 0.21\\
\bottomrule
    \end{tabular}
    \caption{AUC and misclassification rate. The results for qDC, qKSVM, and sqKSVM are from  Table 2 of \cite{moradi2022clinical}, where the two lines for each method represent two different approaches to embedding the 8 features selected from the entire. We fill NA under the Error rate since only AUC values are reported in that table. In contrast, QNN-A embeds all $12$ features via Sinusoidal-friendly Embedding, where the first row corresponds to the URRM resampling, and the second row corresponds to the SMOTE resampling.}
    \label{model performance}
\end{table}

\section{Conclusion}
We have proposed four strategies to enhance the QNN's expressibility. Compared with Arcsin embedding, the proposed Sinusoidal-friendly embedding performs better in fitting Sinusoidal-based nonlinear functions. We also propose to use hybrid embedding to enable the QNN to express general nonlinear functions, including sinusoidal and non-sinusoidal terms. 
The second strategy, redundant measurement, does not increase the number of different basis functions but introduces more parameters to basis functions' coefficients, which makes the QNN's training easier. 
The third strategy, the post-measurement function, creates more different basis functions, thereby increasing the QNN's expressibility without using additional quantum circuit gates and qubits. Our analysis shows that \textit{as the number of qubits increases, the number of basis functions grows exponentially} (see Remark 1), which indicates that QNNs outperform classical NNs in terms of expressibility. 
The last strategy, random training data, helps avoid getting stuck in local optima during training. 
These strategies can benefit both QNNs and VQAs. The embedding and post-measurement function should leverage problem-specific knowledge. 

In the future, we plan to further investigate each strategy, e.g., repeated embedding, different types of measurement, and a better understanding of the impacts of random training data, and to apply our QNN to classification and regression problems in different domains (such as AC power systems, see Remark 3), based on problem-specific knowledge. 

\bibliographystyle{quantum}
\bibliography{mybibliography} 
\section{Appendix}

\subsection{Measure the First Qubit of a 2-qubit Circuit}
\label{section_appen_measure_1st_qubit_of_2qubit_circuit}
This section analyzes the measurement at the first qubit of a 2-qubit circuit (depicted in Figure \ref{figure_3_circuits_using_different_embedding}b), which uses the Sinusoidal-friendly embedding. 
We divide the rest of this section into four parts: input state, output state, measurement expectation, and basis functions.

\subsubsection{Input State.}
Eqs. (\ref{equ_input_state2})$\sim$(\ref{equ_input_state2_3}) provide the quantum input state ($|\psi_{in}\rangle_2$) in a vector form, where the subscript 2 indicates a 2-qubit circuit:
\begin{eqnarray}
\label{equ_input_state2}
    |\psi_{in}\rangle_2 &=&R_y (x_1) |0 \rangle\otimes  R_y (x_2) |0\rangle\\
    &=&\left[\begin{array}{c}
     \cos \frac{x_1}{2}\\
     \sin \frac{x_1}{2}
   \end{array}\right]\otimes \left[\begin{array}{c}
     \cos \frac{x_2}{2}\\
     \sin \frac{x_2}{2}
   \end{array}\right] \label{equ_input_state2_2} \\
    &=& \left[\begin{array}{c}
     \cos \frac{x_1}{2} \cos \frac{x_2}{2}\\
     \cos \frac{x_1}{2} \sin \frac{x_2}{2}\\
     \sin \frac{x_1}{2} \cos \frac{x_2}{2}\\
     \sin \frac{x_1}{2} \sin \frac{x_2}{2}  \label{equ_input_state2_3}
   \end{array}\right]
\end{eqnarray}

\subsubsection{Output State.}
Eqs. (\ref{equ_output_state2})$\sim$(\ref{equ_output_state2_3}) give the output state ($| \psi_{out} \rangle_2$, where the subscript 2 indicates a 2-qubit circuit), in a vector form:
\begin{eqnarray}
\label{equ_output_state2}
 |\psi_{out} \rangle_2 &=& U_2 |\psi_{in} \rangle_2\\
  &=& 
  \left[\begin{array}{cccc}
    u_{11} & u_{12} & u_{13} & u_{14}\\
    u_{21} & u_{22} & u_{23} & u_{24}\\
    u_{31} & u_{32} & u_{33} & u_{34}\\
    u_{41} & u_{42} & u_{43} & u_{44}
  \end{array}\right]      
  \left[\begin{array}{c}
    \cos \frac{x_1}{2} \cos \frac{x_2}{2}\\
    \cos \frac{x_1}{2} \sin \frac{x_2}{2}\\
    \sin \frac{x_1}{2} \cos \frac{x_2}{2}\\
    \sin \frac{x_1}{2} \sin \frac{x_2}{2}
  \end{array}\right]  \label{equ_output_state2_2}\\
  &=&\begin{bmatrix}
  \langle \mathbf{u}_1|\psi_{in}\rangle_2\\
  \langle \mathbf{u}_2|\psi_{in}\rangle_2\\
  \langle \mathbf{u}_3|\psi_{in}\rangle_2\\
  \langle \mathbf{u}_4|\psi_{in}\rangle_2
  \end{bmatrix} \label{equ_output_state2_3}
\end{eqnarray}
where $U_2$ is a $4\times 4$ unitary matrix representing the quantum circuit shown in Figure \ref{figure_3_circuits_using_different_embedding}b, and 
$\mathbf{u}_i=\begin{bmatrix}u_{i1}&u_{i2}&u_{i3}&u_{i4}\end{bmatrix}^T$ with $i=1,2,3,4$.

To better understand $U_2$, we provide a concrete example here. Assuming $U_2$ represents a 2-qubit quantum circuit that consists of an $R_y$ gate at each qubit, we can write $U_2$ as:
\begin{eqnarray}
    U_2&=&\left[\begin{array}{cccc}
            u_{11} & u_{12} & u_{13} & u_{14}\\
            u_{21} & u_{22} & u_{23} & u_{24}\\
            u_{31} & u_{32} & u_{33} & u_{34}\\
            u_{41} & u_{42} & u_{43} & u_{44}
      \end{array}\right]      \label{equ_U2_example_1} \\ 
    &=&R_y^1(\theta_1) \otimes R_y^2(\theta_2)  \label{equ_U2_example_2} \\
     &=&\left[\begin{array}{cc}
              \cos \frac{\theta_1}{2} & - \sin \frac{\theta_1}{2}\\
              \sin \frac{\theta_1}{2} &   \cos \frac{\theta_1}{2}   
        \end{array}\right]       \otimes 
        \left[\begin{array}{cc}
              \cos \frac{\theta_2}{2} & - \sin \frac{\theta_2}{2}\\
              \sin \frac{\theta_2}{2} &   \cos \frac{\theta_2}{2}   
        \end{array}\right]  \label{equ_U2_example_3}
\end{eqnarray}
where superscripts 1 and 2 denote qubits 1 and 2, respectively.
The tensor product result of Eq. (\ref{equ_U2_example_3}) is a $4\times 4$ matrix. Comparing Eq. (\ref{equ_U2_example_1}) with Eq. (\ref{equ_U2_example_3}), we can obtain the expression of $u_{ij}$ for $i,j\in \{1,2,3,4\}$, e.g., 
\begin{align}
u_{11}&=\cos \frac{\theta_1}{2}  \cos \frac{\theta_2}{2}\\
u_{12}&=-\cos \frac{\theta_1}{2}  \sin \frac{\theta_2}{2}\\
u_{43}&=\cos \frac{\theta_1}{2}  \sin \frac{\theta_2}{2}\\
u_{44}&=\cos \frac{\theta_1}{2}  \cos \frac{\theta_2}{2}
\end{align}
That is, each $u_{ij}$ is the product of two sinusoids.
\subsubsection{Measurement Expectation.}
Using Pauli $Z$ to measure the first qubit, we obtain $\langle Z_0\rangle_2$ which is given in (\ref{equ_Z02_expansion}) and also repeated here where subscript 0 represents a measurement at the first qubit:
\begin{align}
  \langle Z_0\rangle_2 &=\langle \psi_{out} |Z \otimes I|\psi_{out} \rangle_2\nonumber\\
  &=\begin{bmatrix}
      \langle \mathbf{u}_1|\psi_{in}\rangle_2\\
      \langle \mathbf{u}_2|\psi_{in}\rangle_2\\
      \langle \mathbf{u}_3|\psi_{in}\rangle_2\\
      \langle \mathbf{u}_4|\psi_{in}\rangle_2
  \end{bmatrix}^{\dagger}
  \begin{bmatrix}
    1 & 0 &  0 & 0\\
    0 & 1 &  0 & 0\\
    0 & 0 & -1 & 0\\
    0 & 0 &  0 & - 1
  \end{bmatrix}
  \begin{bmatrix}
     \langle \mathbf{u}_1|\psi_{in}\rangle_2\\
     \langle \mathbf{u}_2|\psi_{in}\rangle_2\\
     \langle \mathbf{u}_3|\psi_{in}\rangle_2\\
     \langle \mathbf{u}_4|\psi_{in}\rangle_2
  \end{bmatrix}\nonumber\\
 &=\langle\mathbf{u}_1|\psi_{in}\rangle_2^* \langle\mathbf{u}_1|\psi_{in}\rangle_2
   +\langle\mathbf{u}_2|\psi_{in}\rangle_2^* \langle\mathbf{u}_2|\psi_{in}\rangle_2 \nonumber\\
 &\quad\quad-\langle\mathbf{u}_3|\psi_{in}\rangle_2^* \langle\mathbf{u}_3|\psi_{in}\rangle_2
   -\langle\mathbf{u}_4|\psi_{in}\rangle_2^* \langle\mathbf{u}_4|\psi_{in}\rangle_2 
\end{align}
where the superscripts $\dagger$ and $*$ denote the conjugate transpose and complex conjugate, respectively, and 
$Z$ is the Pauli-Z operator: \begin{align}
Z=\begin{bmatrix}1&0\\0&-1\end{bmatrix}\end{align}
Expanding the first term of Eq. (\ref{equ_Z02_expansion}), we get:
\begin{eqnarray}
\label{equ_u1_psiin_u1_psiin}
&&\langle\mathbf{u}_1|\psi_{in}\rangle_2^*      \langle\mathbf{u}_1|\psi_{in}\rangle_2\nonumber \\
&&=u_{11}^{\ast} u_{11} \cos^2 \frac{x_1}{2} \cos^2 \frac{x_2}{2}     + u_{11}^{\ast} u_{12} \cos^2\frac{x_1}{2}\cdot\cos \frac{x_2}{2}\sin \frac{x_2}{2}+\nonumber \\&\quad&
       u_{11}^{\ast} u_{13} \sin \frac{x_1}{2} \cos \frac{x_1}{2}  \cos^2 \frac{x_2}{2}      +u_{11}^{\ast} u_{14} \cos \frac{x_1}{2} \cos \frac{x_2}{2} \sin  \frac{x_1}{2} \sin \frac{x_2}{2}  \nonumber\\&&+u_{12}^{\ast} u_{11}\cdot\cos^2 \frac{x_1}{2}\sin \frac{x_2}{2} \cos \frac{x_2}{2} 
      +u_{12}^{\ast} u_{12} \cos^2  \frac{x_1}{2} \sin^2 \frac{x_2}{2}+                               \nonumber \\
  &&u_{12}^{\ast} u_{13} \cos \frac{x_1}{2} \sin \frac{x_2}{2} \sin \frac{x_1}{2} \cos \frac{x_2}{2}      
      +u_{12}^{\ast} u_{14}\cdot\cos \frac{x_1}{2}  \sin \frac{x_1}{2} \sin^2 \frac{x_2}{2}\nonumber \\
  &&+u_{13}^{\ast} u_{11} \sin\frac{x_1}{2} \cos \frac{x_1}{2}\cdot\cos^2\frac{x_2}{2}+u_{13}^{\ast} u_{12} \sin \frac{x_1}{2} \cos \frac{x_1}{2} \cos  \frac{x_2}{2} \sin \frac{x_2}{2} \nonumber \\
  &&+u_{13}^{\ast} u_{13} \sin^2 \frac{x_1}{2}  \cos^2 \frac{x_2}{2}      
      +u_{13}^{\ast} u_{14} \sin^2 \frac{x_1}{2} \sin\frac{x_2}{2}\cos\frac{x_2}{2}\nonumber \\
  &&+u_{14}^{\ast} u_{11} \sin \frac{x_1}{2} \cos \frac{x_1}{2} \sin  \frac{x_2}{2} \cos \frac{x_2}{2}+u_{14}^{\ast} u_{12} \sin \frac{x_1}{2}  \cos \frac{x_1}{2} \sin^2 \frac{x_2}{2}\nonumber \\
  &&+u_{14}^{\ast} u_{13} \sin^2  \frac{x_1}{2}\cdot        \sin \frac{x_2}{2} \cos \frac{x_2}{2}+u_{14}^{\ast} u_{14} \sin^2 \frac{x_1}{2} \sin^2 \frac{x_2}{2}
\end{eqnarray}

\subsubsection{Basis Functions.}
Regardless of the coefficients, there are 9 distinct terms on the right-hand side of Eq. (\ref{equ_u1_psiin_u1_psiin}), which are:
\begin{align}
\label{equ_u1_psiin_9_basis_functions}
\begin{split}
  &\left\{ \cos^2 \frac{x_1}{2} \cos^2 \frac{x_2}{2},   
     \cos^2\frac{x_1}{2} \cos\frac{x_2}{2}\sin \frac{x_2}{2}, \sin \frac{x_1}{2} \cos \frac{x_1}{2} \cos^2 \frac{x_2}{2},\right.\\
     &\quad\cos \frac{x_1}{2} \cos \frac{x_2}{2} \sin  \frac{x_1}{2} \sin \frac{x_2}{2},\cos^2  \frac{x_1}{2} \sin^2 \frac{x_2}{2}, \cos \frac{x_1}{2}  \sin \frac{x_1}{2} \sin^2 \frac{x_2}{2},\\
     &\left.\quad\sin^2 \frac{x_1}{2} \cos^2 \frac{x_2}{2},\sin^2 \frac{x_1}{2} \sin\frac{x_2}{2} \cos\frac{x_2}{2}, \sin^2 \frac{x_1}{2} \sin^2 \frac{x_2}{2}\right\}
\end{split}
\end{align}
By virtue of the trigonometric identities given in Eqs. (\ref{equ_trig_identity_1})-(\ref{equ_trig_identity_3}), it is not difficult to see that each element in (\ref{equ_u1_psiin_9_basis_functions}) is a linear combination of 9 different basis functions, forming the set $\mathcal{BF}_2$ given in Eq. (\ref{equ_BF2}).
\begin{eqnarray}
      \sin^2 \frac{x_i}{2} = [1-\cos(x_i)]/2  \label{equ_trig_identity_1}
\\    \cos^2 \frac{x_i}{2} = [1+\cos(x_i)]/2  \label{equ_trig_identity_2}
\\    \sin \frac{x_i}{2} \cos \frac{x_i}{2} = \sin(x_i)/2  \label{equ_trig_identity_3}
\end{eqnarray}
\begin{equation}
        \mathcal{BF}_2 = \{1, \sin x_1, \sin x_2, \cos x_1, \cos x_2, \cos x_1\sin x_2, \sin x_1\cos x_2, \cos x_1\cos x_2, \sin x_1\sin x_2\}\nonumber
\end{equation}
\textit{In summary, using Pauli-Z to measure the first qubit of the 2-qubit circuit (given in Figure \ref{figure_3_circuits_using_different_embedding}b), we obtain 9 different basis functions}, i.e., $\mathcal{BF}_2$.

\subsection{Calculate the Number of Different Basis Functions for an n-qubit Circuit}
\label{section_appen_number_of_basis_functions}

\begin{figure*}[t]
    \centering
    \includegraphics[width=\textwidth]{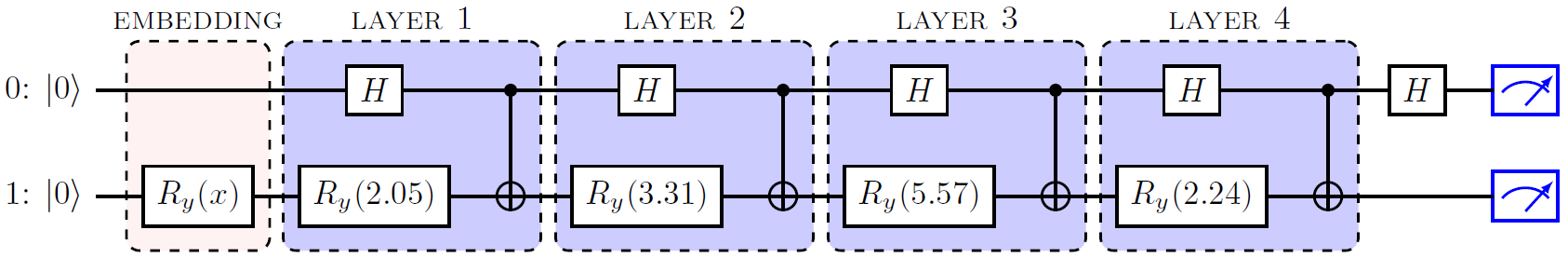}
    \caption{Quantum circuit used by the QNN to fit $f_{1v3}$.}
    \label{figure_appendix_quantum_circuit_f1v3}
\end{figure*}

\begin{figure*}[t]
    \centering
    \includegraphics[width=\textwidth]{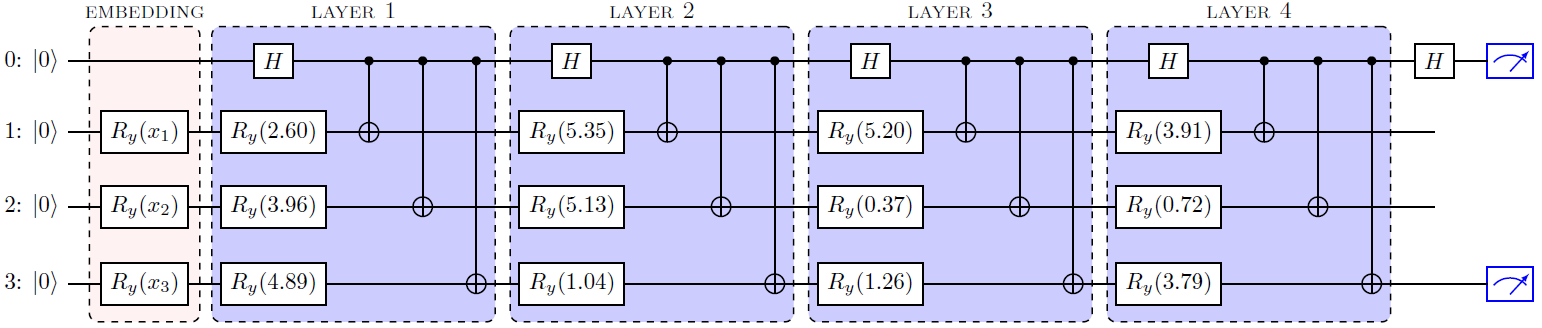}
    \caption{Quantum circuit used by the QNN to fit $f_3$.}
    \label{figure_appendix_quantum_circuit_f3}
\end{figure*}

\begin{figure}[t]
    \centering
    \begin{subfigure}[b]{0.4\textwidth}
    \centering
    \includegraphics[width=0.9\textwidth]{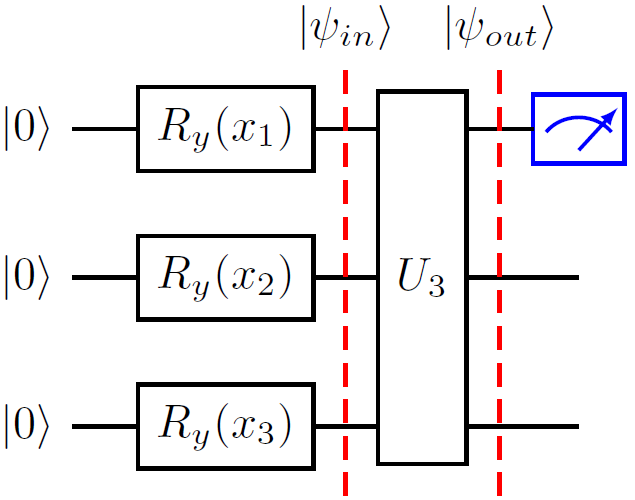}
    \caption{3-qubit circuit}
    \end{subfigure}
    \begin{subfigure}[b]{0.4\textwidth}
    \centering
    \includegraphics[width=0.9\textwidth]{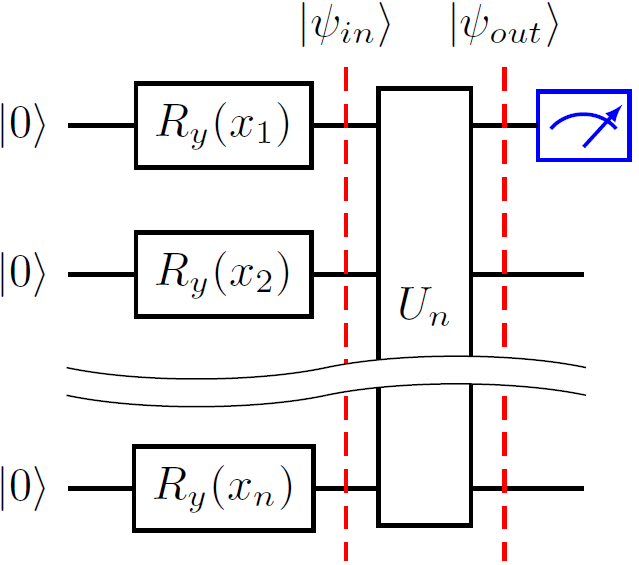}
    \caption{$n$-qubit circuit}
    \end{subfigure}
\caption{Quantum circuits using the Sinusoidal-friendly embedding. Subfigure b uses two waves to represent the initial states, $R_y$ gates, and output of the 3rd$\sim (n-1)$th qubits.}
\label{figure_2_circuits_3qubits_nqubits}
\end{figure}

This section calculates the number of different basis functions for a $n$-qubit quantum circuit, which is shown in Figure \ref{figure_2_circuits_3qubits_nqubits}b. We divide the rest of this section into four parts: input state, output state, measurement expectation, and basis functions.

\subsubsection{Input State.}
We start our analysis with a 3-qubit circuit (shown in Figure \ref{figure_2_circuits_3qubits_nqubits}a).
Eq. (\ref{equ_input_state3}) gives the quantum input state $|\psi_{in}\rangle_3$ of the 3-qubit circuit. In the rest of this section, the number at the subscripts of $|\psi_{in}\rangle$ and $|\psi_{out}\rangle$ denotes the number of qubits in the circuit.
\begin{eqnarray}
\label{equ_input_state3}
    |\psi_{in}\rangle_3 &=&  R_y (x_1) |0 \rangle   
                   \otimes   R_y (x_2) |0   
                   \otimes   R_y (x_3) |0\rangle \nonumber\\
    &=&
    \left[
        \begin{array}{c}
          \cos \frac{x_1}{2}\\
          \sin \frac{x_1}{2}
        \end{array}
   \right]    \otimes   
   \left[
       \begin{array}{c}
         \cos \frac{x_2}{2}\\
         \sin \frac{x_2}{2}
       \end{array}
   \right]    \otimes   
   \left[
       \begin{array}{c}
         \cos \frac{x_3}{2}\\
         \sin \frac{x_3}{2}
       \end{array}
   \right] \nonumber \label{equ_input_state3_2}\\
    &=& 
    \left[
        \begin{array}{c}
             \cos \frac{x_1}{2}  \cos \frac{x_2}{2}  \cos \frac{x_3}{2}\\
             \cos \frac{x_1}{2}  \cos \frac{x_2}{2}  \sin \frac{x_3}{2}\\
             \cos \frac{x_1}{2}  \sin \frac{x_2}{2}  \cos \frac{x_3}{2}\\
             \cos \frac{x_1}{2}  \sin \frac{x_2}{2}  \sin \frac{x_3}{2}\\
             \sin \frac{x_1}{2}  \cos \frac{x_2}{2}  \cos \frac{x_3}{2}\\
             \sin \frac{x_1}{2}  \cos \frac{x_2}{2}  \sin \frac{x_3}{2}\\
             \sin \frac{x_1}{2}  \sin \frac{x_2}{2}  \cos \frac{x_3}{2}\\ 
             \sin \frac{x_1}{2}  \sin \frac{x_2}{2}  \sin \frac{x_3}{2} \label{equ_input_state3_3}
       \end{array}
   \right]
\end{eqnarray}

Let $|\psi_{in}\rangle_n$ be the input state of an $n$-qubit quantum circuit (shown in Figure \ref{figure_2_circuits_3qubits_nqubits}b). 
For the convenience of analysis, we use a general expression to represent any element in $|\psi_{in}\rangle_n$, as shown in Eq. (\ref{equ_s1_s2_si_sn}):
\begin{equation}
\label{equ_s1_s2_si_sn}
    s_1 \times s_2 \times \cdots \times s_i \times \cdots \times s_n    
\end{equation}
where $s_i \in \{\sin \frac{x_i}{2}, \cos \frac{x_i}{2}\}$ for $i=1,2,\cdots,n$. Eqs. (\ref{equ_input_state2_3}) and Eq. (\ref{equ_input_state3_3}) can be used to verify Eq. (\ref{equ_s1_s2_si_sn}) in the case of $n=2$ and $n=3$, respectively. 
For example, for $n=3$, the first element in Eq. (\ref{equ_input_state3_3}) is $\cos \frac{x_1}{2}  \cos \frac{x_2}{2}  \cos \frac{x_3}{2}$, which can be expressed as $s_1 \times s_2 \times s_3$ with 
$s_1= \cos \frac{x_1}{2}$, 
$s_2= \cos \frac{x_2}{2}$, and
$s_3= \cos \frac{x_3}{2}$.

Based on permutation, we can easily know from Eq. (\ref{equ_s1_s2_si_sn}) that $|\psi_{in}\rangle_n$ has $2^n$ elements because $s_i$ ($i=1,2,\cdots,n$) each have two different types, $\sin \frac{x_i}{2}$ and $\cos \frac{x_i}{2}$.

\subsubsection{Output State.}
Similar to Eqs. 
(\ref{equ_output_state2})$\sim$
(\ref{equ_output_state2_3}), Eq. 
(\ref{equ_output_state_n}) provides the output state ($| \psi_{out} \rangle_n$) in a vector form:
\begin{equation}
\label{equ_output_state_n}
| \psi_{out} \rangle_n = U_n| \psi_{in} \rangle_n
  =\begin{bmatrix}
      \langle \mathbf{u}^n_1|\psi_{in}\rangle_n\\
      \langle \mathbf{u}^n_2|\psi_{in}\rangle_n\\
      \vdots\\
      \langle \mathbf{u}^n_i|\psi_{in}\rangle_n\\
      \vdots\\
      \langle \mathbf{u}^n_N|\psi_{in}\rangle_n
  \end{bmatrix}
\end{equation}
where $U_n$ is a unitary matrix representing the $n$-qubit quantum circuit shown in Figure \ref{figure_2_circuits_3qubits_nqubits}b. Note that $U_n$ is a $N\times N$ matrix with $N=2^n$.  The $\mathbf{u}^n_i$ represents the transpose of the $i$th row of matrix $U_n$ for $i=1,2,\cdots, N$, and superscript $n$ in $\mathbf{u}^n_i$ indicates that it is for an $n$-qubit circuit.

\subsubsection{Measurement Expectation.}
Similar to Eq. (\ref{equ_Z02_expansion}), using Pauli $Z$ to measure the first qubit of the $n$-qubit circuit, we obtain:
\begin{eqnarray}
  \langle Z_0\rangle_n &=& \langle \psi_{out} |Z \otimes I^{\otimes n-1}|    \psi_{out} \rangle_n  \nonumber \\
 &=&\sum_{i=1}^{2^{n-1}}       \langle\mathbf{u}^n_i|\psi_{in}\rangle_n^* 
\langle\mathbf{u}^n_i|\psi_{in}\rangle_n-\sum_{i=2^{n-1}+1}^{2^{n}} \langle\mathbf{u}^n_i|\psi_{in}\rangle_n^* 
\langle\mathbf{u}^n_i|\psi_{in}\rangle_n  \label{equ_Z0n_expansion}
\end{eqnarray}
where $I^{\otimes(n-1)}$ represents the tensor product of $n-1$ identity matrices. 

\subsubsection{Basis Functions.}
Similar to the analysis given in Eqs. (\ref{equ_Z02_expansion}) and (\ref{equ_u1_psiin_u1_psiin}), we can expand the first term of Eq. (\ref{equ_Z0n_expansion}) to obtain the basis functions of the $n$-qubit circuit. 
Utilizing the expression given in Eq. (\ref{equ_s1_s2_si_sn}), we can use Eq. (\ref{equ_s1_s2_si_sn_product}) as a general expression for any \textit{\textbf{basis function}} obtained by measuring the first qubit of the $n$-qubit circuit:
\begin{equation}
\label{equ_s1_s2_si_sn_product}
    s_1 \times s_2 \times \cdots \times s_i \times \cdots \times s_n  \   \times\   s_1 \times s_2 \times \cdots \times s_i \times \cdots \times s_n    
\end{equation}
where $s_i\in\left\{\sin\frac{x_i}{2}, \cos\frac{x_i}{2}\right\}$ for $i=1,2,\cdots,n$.

By grouping the same terms together, we can reform Eq. (\ref{equ_s1_s2_si_sn_product}) as:
\begin{equation}
\label{equ_s1_s2_si_sn_product_reformed}
    (s_1\times s_1)\times (s_2 \times s_2) \times \cdots \times (s_i\times s_i) \times \cdots \times (s_n\times s_n)
\end{equation}
where $(s_i\times s_i)$, $i=1,2,\cdots,n$, can only be in three types: 
$\sin \frac{x_i}{2}\times \sin \frac{x_i}{2}$, 
$\sin \frac{x_i}{2}\times \cos \frac{x_i}{2}$, and
$\cos \frac{x_i}{2}\times \cos \frac{x_i}{2}$. 
Using Eqs. (\ref{equ_trig_identity_1})-(\ref{equ_trig_identity_3}), we can reform these three types into another three basis functions, i.e., 1, $\sin(x_i)$, and $\cos(x_i)$. Then we can obtain Eq. (\ref{equ_s1_s2_si_sn_product_reformed2}) from Eq. (\ref{equ_s1_s2_si_sn_product_reformed}):
\begin{equation}
\label{equ_s1_s2_si_sn_product_reformed2}
\{1, \sin x_1, \cos x_1\}\times \{1, \sin x_2, \cos x_2\}\times \cdots \{1, \sin x_i, \cos x_i\}\times \cdots \{1, \sin x_n, \cos x_n\}
\end{equation}

The meaning of Eq. (\ref{equ_s1_s2_si_sn_product_reformed2}) is that we can take one element from each of the $n$ sets and multiply the $n$ elements together to \textit{form a basis function}. 
Based on permutation, we know that Eq. (\ref{equ_s1_s2_si_sn_product_reformed2}) can generate $3^n$ different types of basis functions. That is, \textit{measuring the $n$-qubit circuit (Figure \ref{figure_2_circuits_3qubits_nqubits}b) can yield $3^n$ different basis functions}.

Here we provide two examples of Eq. (\ref{equ_s1_s2_si_sn_product_reformed2}). For the case of $n$=1, by setting $n$ to 1 in Eq. (\ref{equ_s1_s2_si_sn_product_reformed2}),
we directly obtain from Eq. (\ref{equ_s1_s2_si_sn_product_reformed2}) that the basis functions is $\{1, \sin(x_1), \cos(x_1)\}$, which is exactly the same as $\mathcal{BF}_S$, given in the paragraph after Eq. (\ref{equ_c0_c1_c2}). 
For the case of $n$=2, by setting $n$ to 2 in Eq. (\ref{equ_s1_s2_si_sn_product_reformed2}), we can easily verify that Eq. (\ref{equ_s1_s2_si_sn_product_reformed2}) yields 9 basis functions which are exactly the same as $\mathcal{BF}_2$, given in Eq. (\ref{equ_BF2}).

\textit{In summary, using Pauli-Z to measure the first qubit of an $n$-qubit circuit can yield $3^n$ different basis functions. That is, the number of different basis functions grows exponentially as the number of qubits increases. \textbf{This indicates that QNNs have exponential superiority over classical NNs in terms of expressibility}.}

\subsection{Quantum Circuit Structure}
\label{section_appen_circuit_figures}

The circuits used by the QNNs are shown in Figures 
\ref{figure_appendix_quantum_circuit_f1v3} and 
\ref{figure_appendix_quantum_circuit_f3}. 

\subsection{Variance Analysis - More Results}
\label{section_appen_variance_analysis}
\begin{figure}[t]
    \centering
    \begin{subfigure}[b]{0.455\textwidth}
    \centering
    \includegraphics[width=\textwidth]{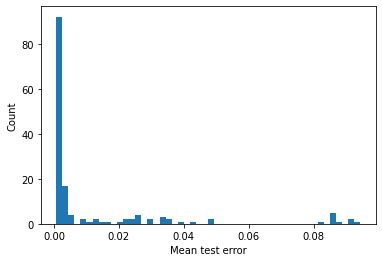}
    \caption{Result for function $f_{1v3}$.}
    \label{figure_hist_f1v3}
    \end{subfigure} 
    \begin{subfigure}[b]{0.455\textwidth}
    \centering
    \includegraphics[width=\textwidth]{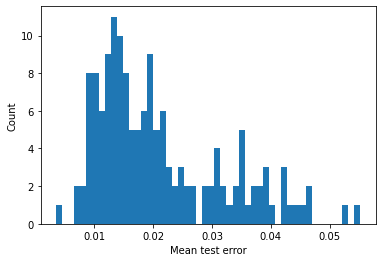}
    \caption{Result for function $f_{1v0}$.}
    \label{figure_hist_f1v0}
    \end{subfigure} 
    \caption{Histogram of mean test error obtained from 150 runs using QNN-A to fit a given function (each run gets a mean test error).}
\end{figure}

This section is a supplement of Section \ref{section_variance_analysis}. 
Figures \ref{figure_hist_f1v3} and \ref{figure_hist_f1v0} show the histogram of mean test error obtained from 150 runs of using QNN-A to approximate functions $f_{1v3}$ and $f_{1v0}$, respectively. Figure \ref{figure_hist_f1v3} shows that most errors are below 0.005 with a few between 0.005 and 0.09. In Figure \ref{figure_hist_f1v0}, most errors are below 0.025, some are between 0.025 and 0.05, and a few go above 0.05.

As mentioned in the main text, the mean value of the 150 values from 150 runs is
0.012 for function $f_{1v3}$ and 
0.021 for function $f_{1v0}$, respectively. 
How good is the test error being around 0.02? 
Note the mean test errors for the results given in Figures \ref{figure_fitting_result_QCL_vs_QNN}d and \ref{figure_fitting_result_QCL_vs_QNN}e are 0.0454 and 0.046, respectively. Considering the fitting performance shown in these two figures looks good, a test error around 0.02 indicates a pretty good fitting performance. 

Therefore, we can conclude that the QNN-A's performance in different runs is very good in most cases and has a very small variance across different runs due to using random training data. To sum up, even with random training data, QNN-A has stable and excellent performance.

\begin{figure}[t]
    \centering
    \includegraphics[width=0.45\textwidth]{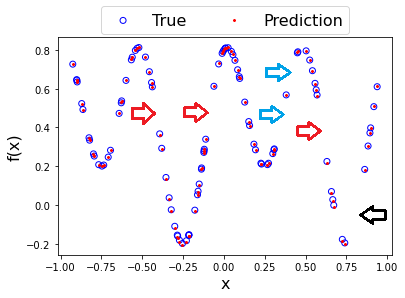}
    \caption{Training data and results of QNN-A in fitting $f_{1v3}$.}
    \label{figure_training_f1v3}
\end{figure}

Figure \ref{figure_training_f1v3} shows the training data and results of QNN-A in fitting $f_{1v3}$, where 100 training data are used. It is interesting to see that the training data have several empty blocks as indicated by arrows in the figure. Note that the empty block near the black arrow is relatively large and near the negative peak of the curve. Also, the two empty blocks near the green arrows are separated by only one training data, that is, this area has almost no training data.\textit{ These empty blocks increase the difficulty of learning the nonlinear function}. 
The fitting result for the corresponding test data is given in Figure \ref{figure_fitting_result_11sub_figures}j. 
Figures \ref{figure_training_f1v3} and \ref{figure_fitting_result_11sub_figures}j show that
QNN-A performs very well in training and test data, respectively. In summary, QNN-A learns the nonlinear function very well even though the training data has relatively large empty blocks.

\subsection{Dataset Description}
\label{section:dataset_description}
The Heart failure clinical records Data Set [42] contains the medical records of 299 patients who had heart failure, collected during their follow-up period, where each patient profile has the following 13 clinical features:
\begin{itemize}
  \item age: age of the patient (years)
  \item anaemia: decrease of red blood cells or hemoglobin (boolean)
  \item high blood pressure: if the patient has hypertension (boolean)
  \item creatinine phosphokinase (CPK): level of the CPK enzyme in the blood (mcg/L)
  \item diabetes: if the patient has diabetes (boolean)
  \item ejection fraction: percentage of blood leaving the heart at each contraction (percentage)
  \item platelets: platelets in the blood (kiloplatelets/mL)
  \item sex: woman or man (binary)
  \item serum creatinine: level of serum creatinine in the blood (mg/dL)
  \item serum sodium: level of serum sodium in the blood (mEq/L)
  \item smoking: if the patient smokes or not (boolean)
  \item time: follow-up period (days)
  \item death event: if the patient deceased during the follow-up period (boolean)
\end{itemize}

\end{document}